\documentclass{aastex62}

\received{}
\revised{}
\accepted{}
\submitjournal{}

\shorttitle{Self-consistent MLCR}
\shortauthors{}

\begin{document}

\title{Self-consistent Color-Mass-to-Light-Ratio relations for Low Surface Brightness Galaxies}

\correspondingauthor{Wei Du}
\email{wdu@nao.cas.cn, wxd165@case.edu}
\author{Wei Du}
\affiliation{Key Laboratory of Optical Astronomy, National Astronomical Observatories, Chinese Academy of Sciences, 20A Datun Road, Chaoyang District, Beijing 100101, China}
\affil{Department of Astronomy, Case Western Reserve University, 10900 Euclid Ave, Cleveland, OH 44106, USA}
\author{Stacy S. McGaugh}
\affil{Department of Astronomy, Case Western Reserve University, 10900 Euclid Ave, Cleveland, OH 44106, USA}



\begin{abstract}
The color $-$ stellar mass-to-light ratio relation (CMLR) is a widely accepted tool 
to estimate the stellar mass (M$_{*}$) of a galaxy. However, an individual CMLR 
tends to give distinct M$_{*}$ for a same galaxy when it is applied in different bands. 
Examining five representative CMLRs from literature, we find that the difference in M$_{*}$ predicted 
in different bands from optical to near-infrared by a CMLR  is  0.1$\sim$0.3 dex.
Therefore, based on a 
sample of low surface brightness galaxies (LSBG) that covers a wide range of 
color and luminosity, we re-calibrated each original CMLR 
in $r$, $i$, $z$, J, H, and K bands to give internally self-consistent M$_{*}$ for a same galaxy.
The $g$-$r$ is the primary color indicator in the re-calibrated relations which show little dependence
on red ($r$ - $z$) or near-infrared (J - K) colors.
Additionally, the external discrepancies in the originally predicted $\gamma_{*}$
by the five independent CMLRs have been greatly reduced after re-calibration, 
especially in near-infrared bands, implying that the near-infrared luminosities are more robust to 
predict $\gamma_{*}$. For each CMLR, the re-calibrated relations provided in this work could produce
internally self-consistent M$_{*}$ from divergent photometric bands, and
are extensions of the re-calibrations from Johnson-Cousin filter system 
by the pioneering work of \citet{McGaugh2014} 
to SDSS filter system.
\end{abstract}



\section{Introduction} \label{sec:intro}
Stellar mass (M$_{*}$) is one of the fundamental physical properties of a galaxy
since it traces star formation and evolution process of the galaxy,
and is crucial to decompose the contributions from stars
and dark matter to the dynamics of a galaxy.
The stellar population synthesis (SPS) technique is
an efficient way to estimate M$_{*}$ of a galaxy,
whereby fitting the SPS models that rely on the extant stellar evolution theory
to galaxy data, either in the form of observed multi-band spectral energy distributions (SEDs),
spectra, or spectral indices of the galaxy. 
Such fit method requires data of SED or spectra,
However, not all the galaxies have multi-band imaging or spectroscopic data,
so a simple color-based method is more practical
to estimate M$_{*}$ of a galaxy.
The pioneering work of \citet[]{Bell2001}(hereafter Bdj01) and \citet[] {Bell2003}(hereafter B03)
have defined relations between color and stellar mass-to-light ratio ($\gamma_{*}$) of galaxies
in the form of equation (1).
\begin{eqnarray}
 {\rm log} \  \gamma_{*}^{j} &=& a_{j} + b_{j}\times {\rm color}
\end{eqnarray} 
 
The $\gamma_{*}$ of a galaxy
can be predicted from the color $-$ stellar mass-to-light ratio relation (CMLR),
and subsequently multiplied by the galaxy luminosity to yield M$_{*}$ of the galaxy.  
The CMLR method requires the minimal data and is hence expedient in all applications related to M$_{*}$ estimate.
Afterwards, a variety of CMLRs have emerged.
A number of these CMLRs are calibrated on model galaxies 
(e.g., \citet[]{Gallazzi2009}, \citet[]{Zibetti2009} (hereafter Z09), 
\citet[]{Into2013}(hereafter IP13), \citet[]{Roediger2015} (hereafter RC15)),
and some are calibrated on samples of observed galaxies,
such as spiral galaxies(e.g., B03,\citet[]{Portinari2004}, \citet[]{Taylor2011}),
dwarf galaxies (e.g., \citet[]{Herrmann2016}),
and low surface brightness galaxies (e.g., \citet[]{Du2020}).
For galaxies the CMLR method could recover $\gamma_{*}$  from a single color 
within an accuracy of $\sim$0.1-0.2 dex \citep{Bell2001},
and could produce equivalent M$_{*}$ to those derived from SED fit method on average \citep{Roediger2015, Du2020}.

However, in the aspect of the CMLR-based M$_{*}$,
\citet[]{McGaugh2014}(hereafter MS14) found the existing CMLR tends to give
different M$_{*}$ for the same galaxy when it is applied in different photometric bands.
Based on a sample of disk galaxies, they re-calibrated several representative CMLRs in Johnson-Cousin filter system
to ultimately produce internally self-consistent M$_{*}$ for the same galaxy 
when it is applied to different bands of $V$, $I$, $K$, and [3.6] bands (with B-V as color indicator).
Inspired by MS14, we expect to extend their work from the Johnson-Cousins bands
to the SDSS optical bands and near-infrared (NIR) bands in this work,
based on a sample of low surface brightness galaxies (LSBGs), 
by first examining the internal self-consistency of a CMLR
in M$_{*}$ estimates from different bands 
and then re-calibrating the CMLR to be able to
give internally self-consistent M$_{*}$ estimates from different bands for the same galaxy.

We describe the data in Section \ref{sec:data} and introduce the five representative CMLR models in Section \ref{sec:SPS}.
We estimated M$_{*}$ from different bands for the sample by the CMLRs,
and internally compared M$_{*}$ from different bands by each individual CMLR, 
and then externally compared M$_{*}$ predicted by different CMLRs in Section \ref{sec:mstar}.
In Section \ref{sec:m2l}, each individual CMLR is re-calibrated to 
be internally self-consistent in M$_{*}$ estimates for the sample, 
when it is applied in different bands from optical to NIR.
We make a discussion in Section \ref{sec:discus},
including the possible second color term to the re-calibrated relations in Section \ref{sec:discus_sub1},
the error budget in $\gamma_{*}$ predicted by the re-calibrated relations in Section \ref{sec:errors},
comparison between originally predicted $\gamma_{*}$ and those predicted by the re-calibrated relations in  Section \ref{sec:discus_sub2},
and comparison between re-calibrated relations in this work and those by MS14 in Section \ref{sec:MS14}. 
A summary and conclusion is given in $\S$~\ref{sec:conclusion}.
Throughout the work, the magnitude is in AB magnitude system,
and the galaxy distance used to calculate the absolute magnitude and luminosity 
is directly from the Arecibo Legacy Fast ALFA Survey (ALFALFA) catalogue \citet{Haynes2018},
which adopts a Hubble constant of $H_{0}$ = 70 km s$^{-1}$ Mpc$^{-1}$.

\section{Data} \label{sec:data}
\subsection {LSBG Sample} \label{sec:sample}
\begin{figure}[ht!]
\centering
\includegraphics[width=0.7\textwidth]{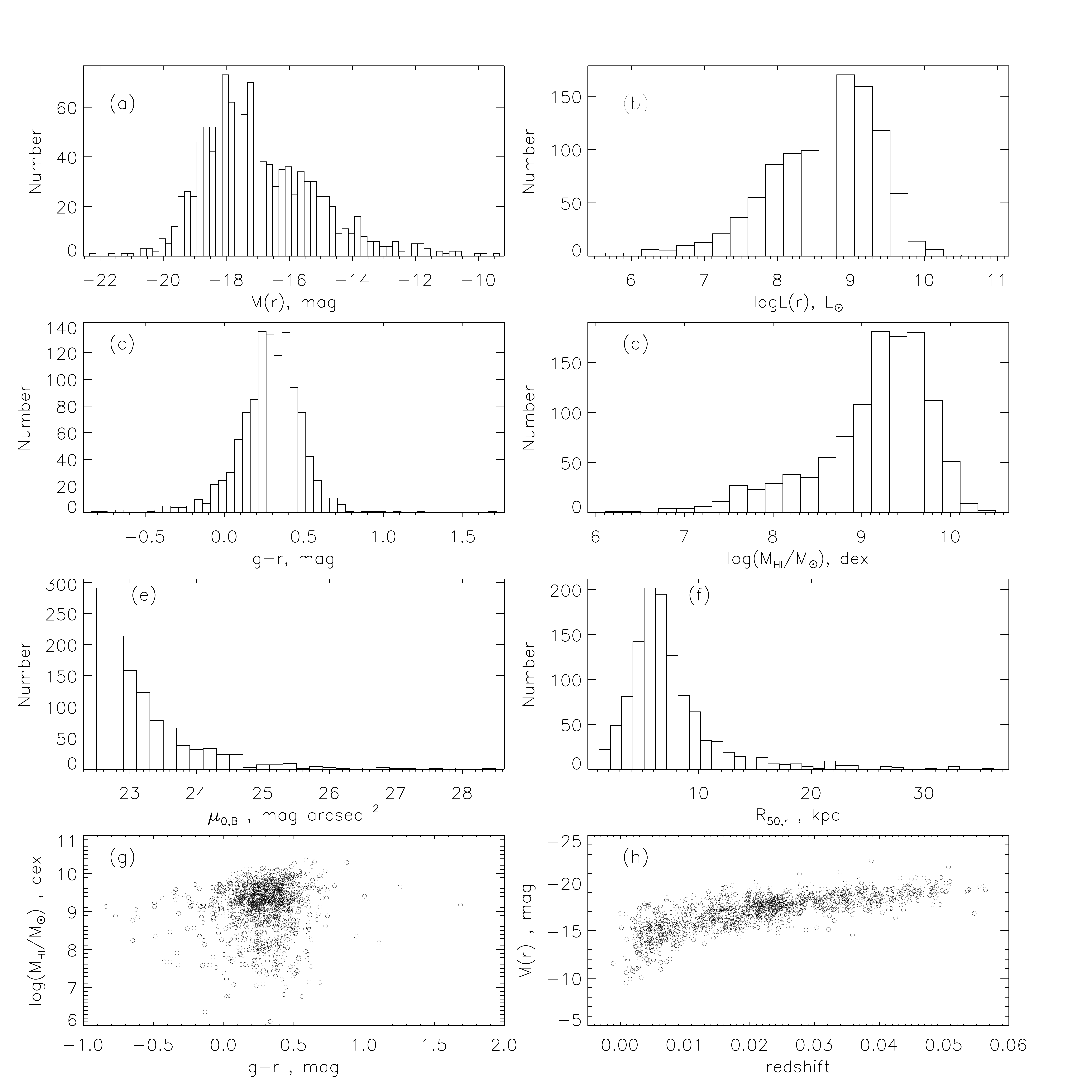} 
\caption{Properties of the LSBG sample. In panels (a) - (f), the distributions of $r$-band absolute magnitude (M(r)), the $r$-band luminosity in logarithm (log L(r)),
 $g$-$r$ color ($g$-$r$), H{\sc{i}} mass in logarithm (log M$_{H_{I}}$/M$_{\odot}$), $B$-band central surface brightness ($\mu_{0,B}$), and effective radius (R$_{50,r}$)
 are shown, respectively.  Panels (g) and (h) show $g$-$r$ versus H{\sc{i}} mass, and M(r) versus redshift, respectively.} \label{fig:property}
\end{figure}

Since low surface brightness galaxies (LSBGs) are typically gas-rich,
we have defined a sample of LSBGs 
from a survey combination of $\alpha$.40 H{\sc{i}} \citep{Haynes2011} 
and SDSS DR7 photometric \citep{Abazajian2009} surveys,
and selection about this sample have been detailedly reported in \citet{Du2015} and \citet{Du2019}.
This sample includes 1129 LSBGs which have B-band central surface brightnesses ($\mu_{0,B}$) 
fainter than 22.5 mag arcsec$^{-2}$ ($\mu_{0,B} >$ 22.5),
and has extended the parameter space
covered by the previous LSBG samples
to fainter luminosity, lower H{\sc{i}} gas mass, and bluer color (Figure \ref{fig:property}).
In color, the full range of this sample is -0.8 $< g-r <$1.7
(the peak at 0.28 and 1$\sigma$ scatter of 0.21),
with 95.4$\%$ within -0.14 $< g-r <$ 0.70
and 68.3$\%$ within 0.07 $< g-r <$ 0.49.
In absolute magnitude, the full range of the sample spans over 10 mag,
with 95.4$\%$ within -13$ <$ M$_{r} <$ -21 mag
and 68.3$\%$ within -15 $ <$ M$_{r} <$ -19 mag.
In terms of luminosity,
it is composed of the dwarf (M$_{B} \geq$-17.0~mag; 54$\%$ of the sample), 
moderate-luminosity (-19.0$<$M$_{B}<$-17.0~mag; 43$\%$),
and giant galaxies (M$_{B} \leq$-19.0~mag; 3$\%$).
In terms of morphology, it is dominated by the late-type spiral and irregular galaxies (Sd/Sm/Im; 84.1$\%$ of the sample),
then the early- and middle-type spiral galaxies (Sa/Sab/Sb/Sbc/Sc/Scd;13.4$\%$), 
and finally the early-type galaxies (E/S0; 0.2$\%$)\citep{Du2019}.
In this work, we intend to re-calibrate several literature CMLRs (Section ~\ref{sec:SPS}) 
based on this sample of LSBGs.

\subsection {Photometry}\label{sec:phot}
The optical images ($griz$ bands) of the sample were downloaded from 
SDSS DR7 \citep[]{Abazajian2009}, and the NIR images (JHK bands) were obtained from UKIDSS \citep{Lawrence2007}.
For each image, we subtracted the sky background, excluded the bright disturbing objects 
around the target galaxy, and replaced the masked pixels with the mean value of the surrounding background pixels. 
The magnitudes of the target galaxy were then measured in these bands in \citet{Du2020}
with SExtractor \citep{Bertin1996} in the dual-image mode,
in which the $r$-band image is regarded as a reference and is used to
detect the galaxy source and define the photometric apertures (center, size and shape). 
Images of the same galaxy in all other bands
are photometrically measured within the same aperture defined in the $r$ band. 
The measured magnitudes in all the bands are corrected for Galactic extinction
using the prescription of \citet{Schlafly2011}. As LSBGs are poor in dust content, 
we do not correct the internal extinction to magnitudes. 
Finally, magnitudes in all the bands were converted to AB magnitude system.
We adopt a distance given in ALFALFA catalogue \citep{Haynes2018}
to compute absolute magnitude and luminosity in each band of $griz$JHK.
As the aperture definition for each galaxy does not vary between wavelength bands,
such measurement gives internally consistent colors.

\begin{table*}\footnotesize
\caption{Original CMLRs based on $g$-$r$ color }
\label{tab:cmlr1}
\begin{center}
\begin{tabular}{lcccccccccccccc}
\hline
\hline
model & IMF & TP-AGB& a$_{r}$ & b$_{r}$ & a$_{i}$ & b$_{i}$ & a$_{z}$ & b$_{z}$ & a$_{\rm J}$ & b$_{\rm J}$ & a$_{\rm H}$ & b$_{\rm H}$ &a$_{\rm K}$ & b$_{\rm K}$\\ 
\hline
\hline
B03& `diet' Salpeter &Girardi&  -0.306& 1.097 & -0.222& 0.864&-0.223& 0.689&-0.172& 0.444&-0.189& 0.266&-0.209& 0.197 \\
IP13 & Kroupa &Marigo&  -0.663&1.530&-0.633& 1.370&-0.665& 1.292&-0.732& 1.139&-0.880& 1.128&-0.945& 1.153 \\
Z09 &Chabrier &Marigo&  -0.840 & 1.654&-0.845& 1.481&-0.914& 1.382&-1.007& 1.225&-1.147& 1.144&-1.257& 1.119\\
RC15(BC03)&Chabrier&Girardi&  -0.792 & 1.629&-0.771& 1.438&-0.796& 1.306& --      & --        &-0.920& 0.980&--        & -- \\
RC15(FSPS)&Chabrier&Marigo&-0.647 &1.497 &-0.602& 1.281&-0.583& 1.102&--       & --       &-0.605& 0.672& --        & --\\
\hline
\hline
\multicolumn{15}{p{1.0\textwidth}}{Notes. Stellar mass-to-light ratios ($\gamma_{*}$) in
SDSS $r$, $i$, $z$ and NIR J, H, K bands are given
by the CMLRs of \citet[B03]{Bell2003}, \citet[IP13]{Into2013}, \citet[Z09]{Zibetti2009},
\citet[]{Roediger2015} based on BC03 model (RC15(BC03)), and 
\citet[]{Roediger2015} based on FSPS model (RC15(FSPS) in the formula of
log $\gamma_{*}^{j}$ = $a_{j}$+$b_{j}\times(g - r)$. For reference, the initial mass function (IMF)
and the TP-AGB prescription adopted by
each CMLR model are also given. 
For IMF,  the `Kroupa' denotes the \citet{Kroupa1998} IMF, and `Chabrier' denotes
the \citet{Chabrier2003} IMF.
For TP-AGB, the `Girardi' denotes the simplified TP-AGB
prescriptions \citep[e.g.][]{Girardi1998,Girardi2000,Girardi2002}, 
while `Marigo' denotes the relatively new TP-AGB prescriptions
\citep[e.g.][]{Marigo2007,Marigo2008},
which incorporate a larger number of TP-AGB stars.}
  \end{tabular}
   \end{center} 
 \end{table*}
 
\begin{figure}[ht!]
\centering
\includegraphics[width=1.0\textwidth]{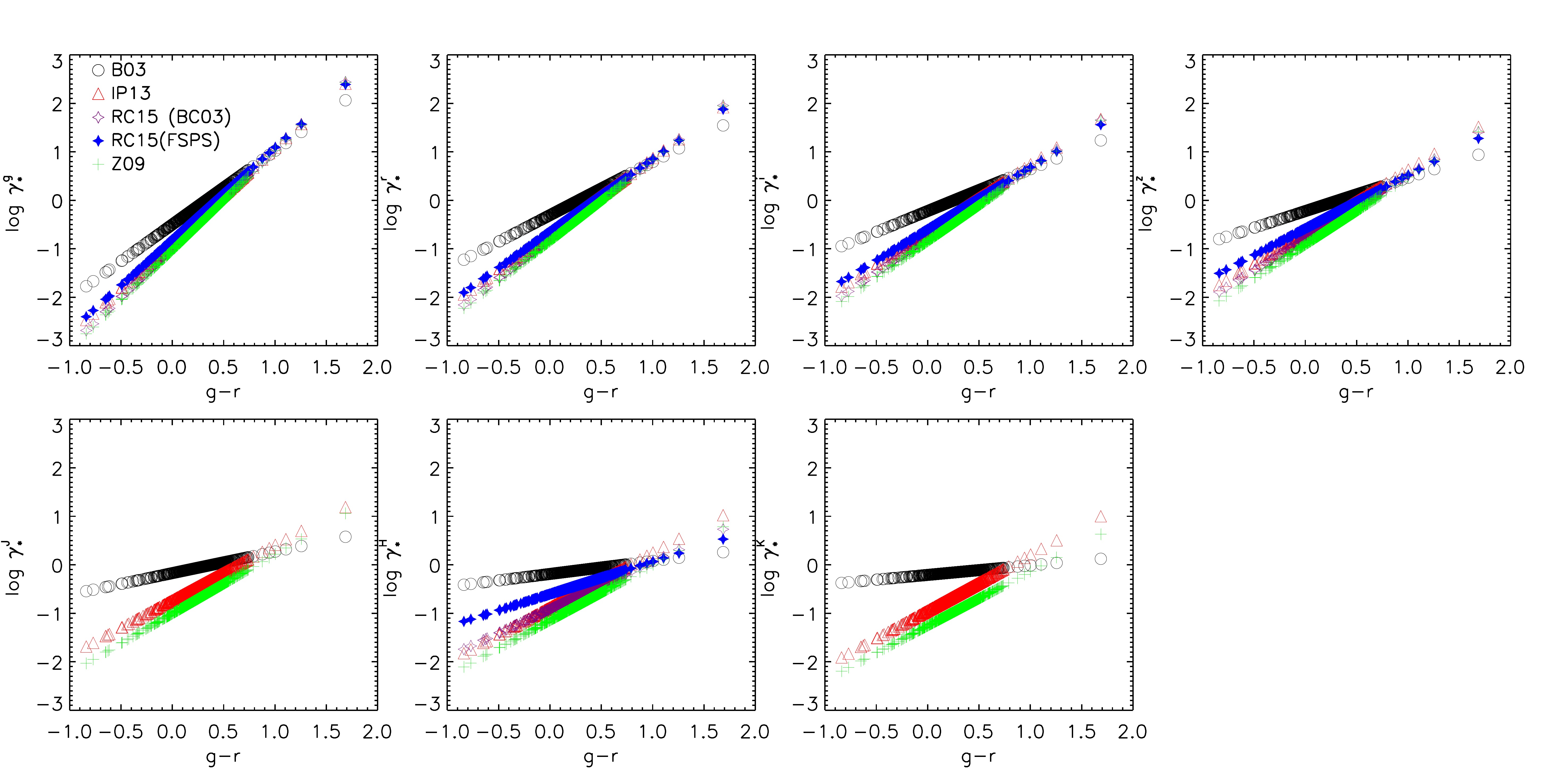} 
\caption{ Relation between $g$ - $r$ color and log $\gamma{*}^{j}$ ($j$= $g$, $r$, $i$, $z$, J, H, and K bands)
from the CMLRs of B03 with an assumption of a `diet' Salpeter IMF (black circles), Z09 with an assumption of a \citet{Chabrier2003} IMF (green plus), 
IP13 with an assumption of a \citet{Kroupa1998} IMF (red triangles), RC15(BC03) with an assumption of \citet{Chabrier2003} IMF (purple open stars), 
and RC15(FSPS) with an assumption of \citet{Chabrier2003} IMF (blue filled stars).}\label{fig:MLCR}
\end{figure}

\section{CMLR Models}\label{sec:SPS}
In the pioneering work of MS14,
the CMLRs of B03, Z09, IP13, and \citet{Portinari2004}(P04)
are re-calibrated in the V, I, K, and [3.6] bands with B - V as the color indicator.
In this work, we aim to extend MS14 
from Johnson-Cousins filters to SDSS optical and two more NIR filters.
Besides the three CMLRs of B03, Z09, and IP13 studied in MS14 which also provide relations in SDSS bands,
two more CMLRs of the RC15 based on the BC03 stellar population model (RC15(BC03)) 
and the  FSPS model (RC15(FSPS)) will be considered.

B03 is an empirical relation while the others (Z09, IP13, RC15) are theoretical.
 B03 is based on a sample of observed galaxies,
 which are mostly bright galaxies (13 $\leq r \leq$17.5~mag)
 with high surface brightnesses (HSB; $\mu_{r} <$21~mag~arcsec$^{-2}$),
 and spans a full range of 0.2 $< g$-$r <$ 1.2 with most galaxies within the range of 0.4 $< g$-$r<$1.0.
 (Figure 5 in B03 paper).
For the theoretical relations, Z09 is based on a 
library of stellar population models from the 2007 version of BC03 (CB07),
which covers from 0 to 20 Gyr in age, 6 values in metallicity (Z=0.0001 to 0.05),
and spans a range of -0.3 $< g$-$i <$ 2.6. 
IP13 is based on a sample of stellar population models from the isochrones of the Padova,
which covers from 0.1 to 12.6 Gyr in age, 7 values in metallicity (Z=0.0001,0.0004,0.001,0.004,0.008,0.019,0.03),
 and spans a range of from 0.25 $< g$-$r< $ 0.75. 
 RC15 is also based on stellar population models from BC03 or FSPS,
 which spans a range of -0.25 $< g$-$r< $1.65 for RC15(BC03) and a range of  -0.1 $< g$-$r< $ 1.65 for RC15(FSPS)
(Figure 7 in RC15).
By comparison, the sample of observed data of LSBGs (Section ~\ref{sec:data}), 
have a range of $\mu_{r} >$21~mag~arcsec$^{-2}$ and $r >$17.5~mag,
and 73$\%$ of the sample is bluer than $g$-$r$=0.4.
In Table ~\ref{tab:cmlr1},  we tabulated these 5 representative CMLRs of B03, IP13, Z09, RC15(BC03), and RC15(FSPS),
in $r$, $i$, $z$, J, H, and K bands with $g$ - $r$ as the color indicator.

Figure \ref{fig:MLCR} presents the stellar mass-to-light ratios ($\gamma_{*}$) in $j$ band ($\gamma_{*}^{j}$, $j$=$g$, $r$, $i$, $z$, J, H, K) 
predicted by each CMLR (Table ~\ref{tab:cmlr1}) for the sample,
showing the beads-on-a-string nature of $\gamma_{*}$ from the single color-based CMLR method.
It uncovers that the CMLR-based method fails to reproduce the intrinsic scatter of $\gamma_{*}$ 
expected from variations in star formation histories (SFH).
In each panel,  $\gamma_{*}$ from different CMLRs differ from each other
due to distinct choices of initial mass function (IMF), star formation history (SFH), 
and stellar evolutionary tracks by different CMLR models.

Different IMFs primarily differ in the treatment of low mass stars.
The IMF that includes a larger number of low mass stars normally
produces a higher $\gamma^{*}$ at a given color than the IMFs
incorporating a smaller number of low mass stars. 
This is in principle because the low mass stars could greatly enhance 
the stellar mass but alter little the luminosity. 
Therefore, diverse IMFs would predominantly lead to
difference in the zero-point of CMLRs.
For example, stellar mass estimates based on
a Chabrier or Salpeter IMF differ by 0.3 dex, with the latter being higher \citep{Roediger2015}.
As listed in Table \ref{tab:cmlr1},
B03 adopts a `diet' Salpeter IMF, which includes more low mass stars
than the Chabrier IMF utilized by RC15 and Z09 CMLRs
and the \citet{Kroupa1998} IMF used by IP13 CMLR,
so B03 gives a higher $\gamma^{*}$ than other CMLRs
at a given color (Figure \ref{fig:MLCR}).

Galaxies are expected to have a wide range of SFHs.
The best-fit stellar mass could be significantly changed by different SFHs, 
in particular whether the SFH is continuous (rising/declining) or bursty.
Any burst of star formation will bias the models towards lower $\gamma_{*}$ values 
than the smooth star formation models at a given color .
The uncertainties of $\gamma_{*}$ in optical due to different SFHs
are $\sim$0.2 dex for quiescent galaxies, 
$\sim$0.3 dex for star-forming galaxies \citep{Kauffmann2003},
$\sim$ 0.5 dex at a given $B$-$R$, and could be up to 0.6 dex in extreme cases \citep{Courteau2014}.
For the CMLRs in this work,
IP13 adopts a single component model of exponential SFH
while other CMLRs in this work are all based on two-component models of SFH.
Z09 and RC15(BC03) both consider the exponentially declining SFHs with a variety of random bursts superimposed.
RC15(FSPS) uses the exponential SFH with only one instantaneous burst added.
B03 assumes the exponential SFH (starting from 12 Gyr in the past) 
with bursts superimposed, but limits the strength of bursts to $\leq$10$\%$ by mass
constrains the burst events to only take place in the last 2 Gyr, so it is relatively smooth.

In Figure ~\ref{fig:MLCR}, 
the discrepancies in $\gamma_{*}$ among the CMLRs in the NIR bands (J, H, and K) 
are obviously larger than the discrepancies in the optical bands ($griz$). 
This primarily rises from the different treatments
of the TP-AGB stars which are the low to intermediate mass stars (0.6 $\sim$ 10 M$_{\odot}$)
in their late life stage, and emit a considerable amount of light in the NIR but little light to the optical.
As listed in Table ~\ref{tab:cmlr1}, B03 and RC15(BC03) adopt a simplified prescription \citep{Girardi2000, Girardi2002} for TP-AGB stars, 
whereas IP13, RC15(FSPS), and Z09 consider a relatively new prescription \citep{Marigo2007,Marigo2008} for TP-AGB stars.
The latter prescription incorporates a larger number of TP-AGB stars, and
would hence greatly enhance the NIR luminosity but alter little to the optical luminosity.
This inevitably results in lower NIR $\gamma_{*}$ but change little to the optical $\gamma_{*}$.

\section{Stellar Mass}\label{sec:mstar}
The average $\gamma_{*}$ in the $u$ band 
suffers more from the perturbations of the young, luminous, blue stars 
which formed recently and radiated a significant amount of light in the blue bands
but contribute little to the galaxy mass. Additionally, the SDSS $u$-band data are of low quality, 
so we shall exclude the $u$-band $\gamma_{*}$ from the following analysis.

For the LSBG sample, we predict $\gamma_{*}^{j}$ ($j$=$g$, $r$, $i$, $z$, J, H, and K bands) by each independent CMLR 
with $g$ - $r$ as the color indicator (Table \ref{tab:cmlr1}), as $g$-$r$ serves as a good color indicator for $\gamma_{*}$.
The predicted $\gamma_{*}^{j}$ are then multiplied by luminosities in $j$ band (Section ~\ref{sec:phot}) 
to produce M$_{*}$ estimates from $j$ band (M$_{*}^{j}$).
We list the mean and the median M$_{*}^{j}$ originally by each CMLR for the sample in the left part in Table ~\ref{tab:mass}. 

We can check external consistency of different CMLRs 
by comparing M$_{*}$ from different CMLRs.
It is apparent that the five CMLRs produce distinct M$_{*}^{j}$ estimates from the $j$ band ($j$=$g$, $r$, $i$, $z$, J, H, and K bands).
In the same $j$ band, B03 gives the highest M$_{*}$ while Z09 yields the lowest M$_{*}$ for the sample.
The difference between M$_{*}$ predicted by B03 and Z09 is 0.3$\sim$0.5 dex in optical bands, and dramatically rises 
up to 0.6$\sim$0.8 dex in NIR bands due to the different treatments for TP-AGB stars (Section \ref{sec:SPS}). 
The external inconsistency is caused by the different choices of the IMF, SFH, and SPS models.

We can examine each CMLR for the internal consistency from different bands.
For any individual CMLR, M$_{*}$ predicted from $g$ band (M$_{*}^{g}$) are closely consistent with M$_{*}$ predicted from $r$ band (M$_{*}^{r}$).
However, M$_{*}^{j}$ ($j$=$i$, $z$, J, H, and K), especially $j$= J, H, and K, deviate from M$_{*}^{r}$ to varying degrees,
and the deviation is progressively increasing as the band goes redder.
\textbf{For instance, the deviation of M$_{*}^{\rm NIR}$ from M$_{*}^{r}$ is 0.1 dex by B03, -0.3 dex by Z09,
and -0.1 $\sim$ -0.3 dex by the three other CMLRs. }
This implies that B03 is nearly internally self-consistent in M$_{*}$ estimate from different bands,
but it has a small tendency to overestimate M$_{*}$ estimates from NIR bands, 
whereas the four other CMLRs all underestimate M$_{*}$ from NIR bands, compared with M$_{*}^{r}$.

\begin{figure}
\centering
\includegraphics[width=1.1\textwidth]{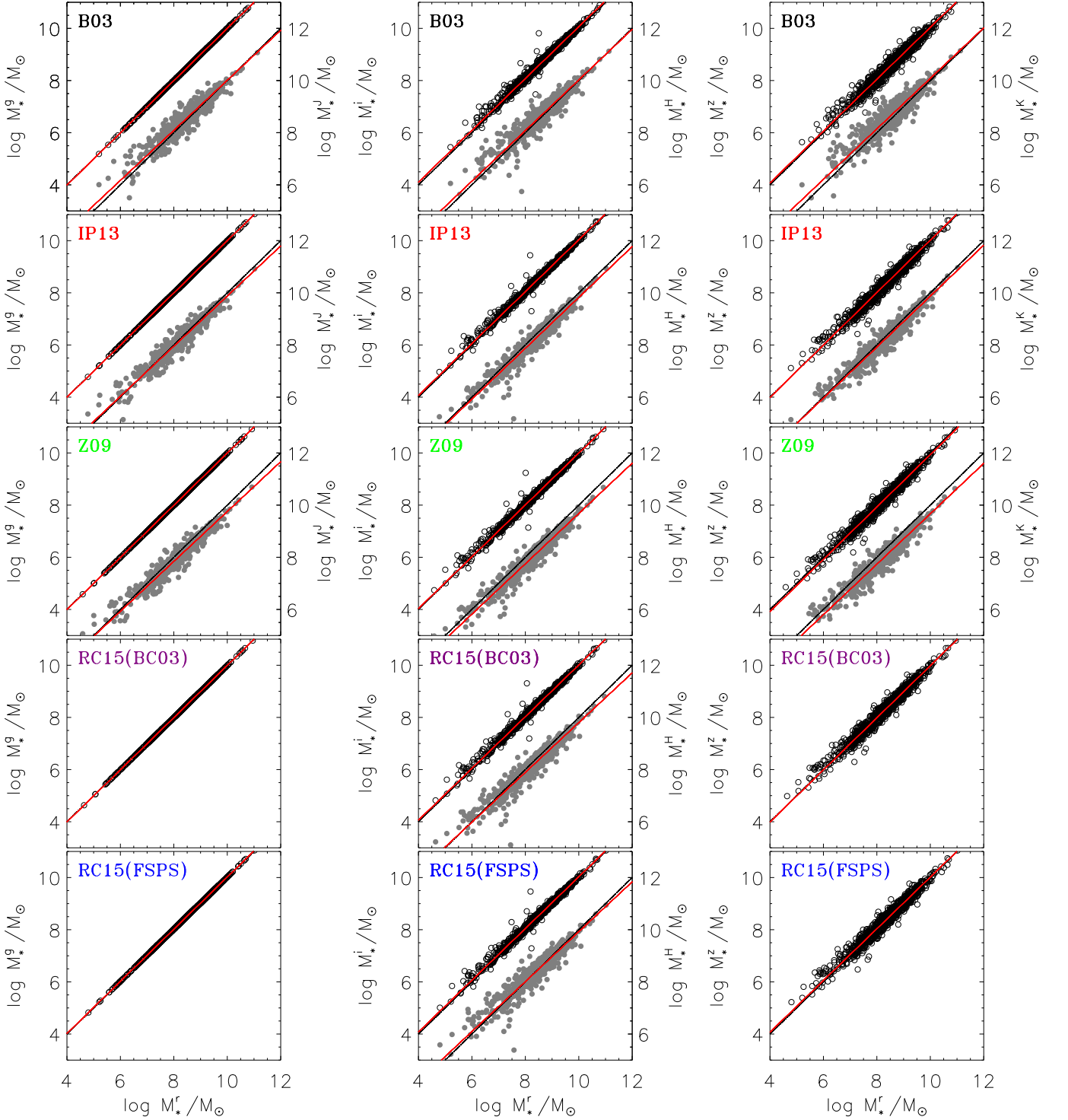} 
\caption{Stellar mass (M$_{*}$) estimates by different CMLRs of B03, IP13, Z09, RC15(BC03), and RC15(FSPS) listed in Table ~\ref{tab:cmlr1}.
For each CMLR, M$_{*}$ estimates from $g$ (open black circles) or J (filled grey circles) bands are, respectively, plotted against M$_{*}$ from $r$ band (M$_{*}^{r}$) in the left panel. 
M$_{*}$ estimates from $i$ (open black circles) or H (filled grey circles) bands are, respectively, plotted against M$_{*}^{r}$ in the middle panel. M$_{*}$ estimates from $z$ (open black circles) or K (filled grey circles) bands are, respectively, plotted against M$_{*}^{r}$ in the right panel. For each panel, the two cases are offset for clarity, and the black dashed lines are the line of unity, and the red solid lines are the fit to the data. If the CMLR were internally self-consistent, the data would follow the line of unity (black dashed line). However, the fit line which the data follow obviously deviates from the line of unity, expect for data of $g$ v.s. $r$ bands. It should be noted that RC15 does not provide relations in J and K bands.}\label{fig:mstar_discussion}
\end{figure}

\begin{figure}
\centering
\includegraphics[width=1.1\textwidth]{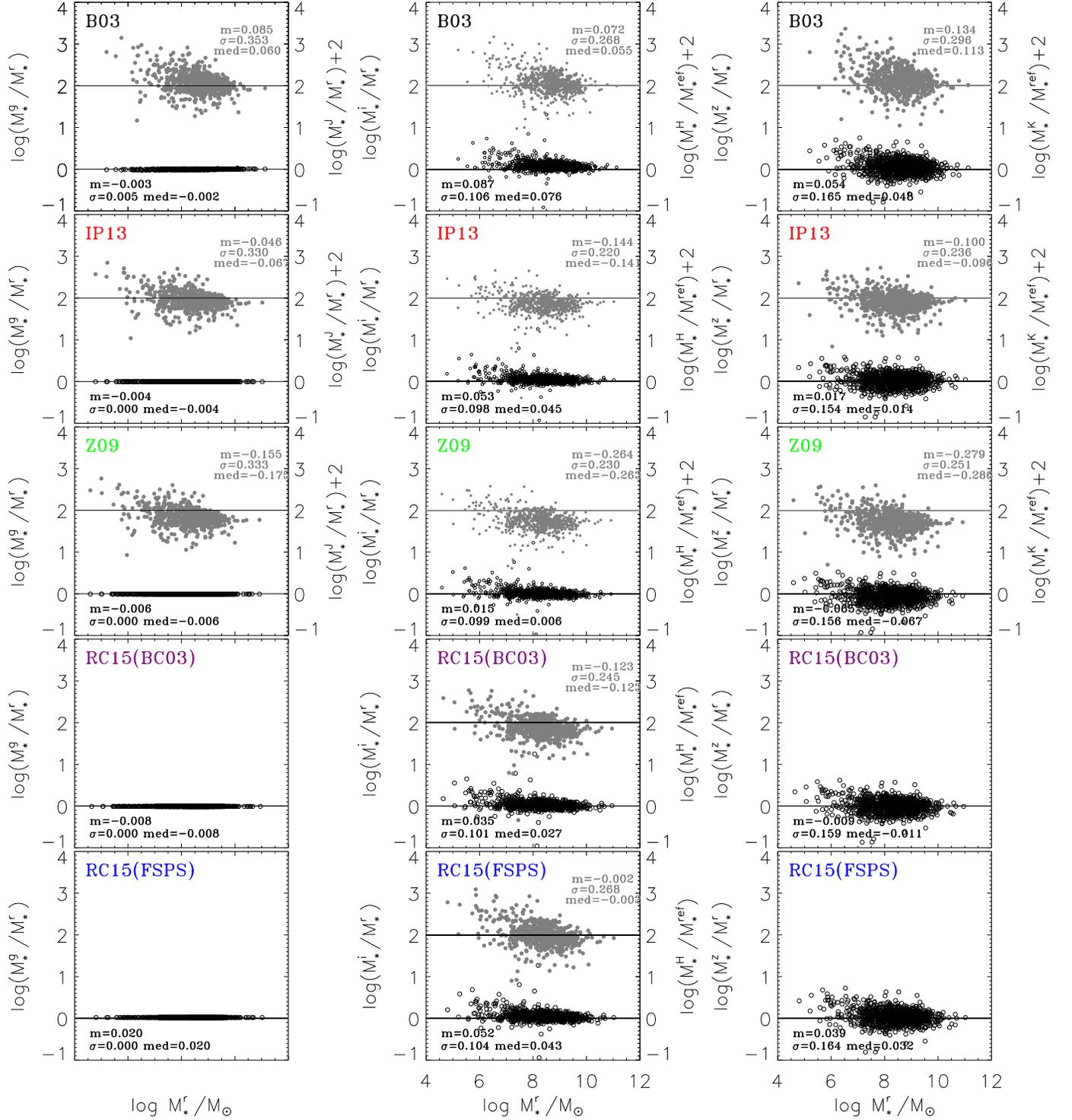} 
\caption{Residuals of data from the line of unity in each panel in Figure ~\ref{fig:mstar_discussion}.
For each CMLR, residuals of the data from the line of unity in the $g$, $i$, and $z$ bands are shown as open black circles in the lower region of 
the left, middle, and right panels, respectively. For clarity, residuals of the data from the line of unity in J, H, and K bands are offset by +2 and 
shown as grey filled circles in the upper region of the left, middle, and right panels, respectively. 
The black and grey solid lines in each panel are the zero-residual lines for the corresponding data. }\label{fig:residual}
\end{figure}

In Figure ~\ref{fig:mstar_discussion}, we show M$_{*}^{j}$ ($j$= $g$, $r$, $i$, $z$, J, H, and K) 
against M$_{*}^{r}$ predicted by each CMLR for the sample (black open circles or grey filled circles). 
For each panel, the black dashed lines represent the line of unity for the data.
If the CMLR is internally self-consistent in M$_{*}$ estimate from band to band, 
the data should exactly follow the line of unity.
However, it does not seem to be the fact
given that the data (black open circles or grey filled circles) in each panel 
obviously deviate from the line of unity (black dashed lines) to different degrees,
except for the data in the panel of M$_{*}^{g}$ versus M$_{*}^{r}$.
This demonstrates that M$_{*}^{g}$ are highly consistent with M$_{*}^{r}$
while M$_{*}^{j}$ ($j$= $i$, $z$, J, H, and K) deviates from M$_{*}^{r}$,
and the deviation is progressively increasing as the band goes redder.
In order to clearly display the deviation of data from the line of unity, 
we plot the residuals of data from the line of unity in Figure ~\ref{fig:residual}.
In the case of internally inconsistency of each CMLR from band to band,
we shall calibrate each CMLR to be internally self-consistent in M$_{*}$ estimates 
from different bands, based on this LSBG sample in $\S$ ~\ref{sec:m2l}.

\begin{table*}
\caption{Mean (upper) and Median (lower) Stellar mass predicted by the original (left part)  and re-calibrated CMLRs (right part) for the LSBG sample.}
\label{tab:mass}
\begin{center}
\begin{tabular}{lccccc|ccccc}
\hline
\hline
band&B03&IP13&R15(BC03)&R15(FSPS)&Z09 &B03&IP13&R15(BC03)&R15(FSPS)&Z09\\
\hline
$g$&8.64&8.43&8.30&8.44&8.26 & 8.64 & 8.43 & 8.30 & 8.44 & 8.26\\
$r$&8.65&8.43&8.31&8.42&8.27  & 8.65 & 8.43 & 8.31 & 8.42 & 8.27\\
$i$&8.73&8.48&8.35&8.47&8.29  & 8.66 & 8.42 & 8.32 & 8.43 & 8.28\\
$z$&8.70&8.44&8.30&8.46&8.21 & 8.66 & 8.42 & 8.32 & 8.43 & 8.28\\
J&8.68&8.30&-- & --&8.06            & 8.62 & 8.38 & -- & -- & 8.24\\
H&8.68&8.21&8.15&8.38&7.97    & 8.62 & 8.37 & 8.28 & 8.39 & 8.23\\
K&8.75&8.26&-- &-- &7.97           & 8.64 & 8.38 & -- & -- & 8.24\\
\hline
$g$&8.75&8.55&8.42&8.56&8.38  & 8.75 & 8.55 & 8.42 & 8.56 & 8.38\\
$r$&8.76&8.55&8.43&8.54&8.39  & 8.76 & 8.55 & 8.43 & 8.54 & 8.39\\
$i$&8.84&8.60&8.46&8.59&8.40  & 8.77 & 8.54 & 8.43 & 8.55 & 8.39\\
$z$&8.82&8.55&8.42&8.58&8.32 & 8.77 & 8.54 & 8.44 & 8.55 & 8.39\\
J&8.83&8.43& --&-- &8.20             & 8.77 & 8.52 & -- & -- & 8.38\\
H&8.80&8.33&8.28&8.51&8.09    & 8.75 & 8.49 & 8.4 & 8.52 & 8.35\\
K&8.88&8.38&-- &-- &8.08           & 8.77 & 8.49 & -- & -- & 8.35\\
\hline
\hline
  \end{tabular}
   \end{center} 
 \end{table*}


\section{Self-consistent M/L-color relations}\label{sec:m2l}
\subsection{Self-Consistent Stellar Masses}
For each individual CMLR, the M$_{*}$ estimates from $g$ band (M$_{*}^{g}$)
closely agree with those from $r$ band (M$_{*}^{r}$) for the sample.
However, the M$_{*}$ estimates from $i$, $z$, J, H, and K bands differ from M$_{*}^{r}$ for the sample 
to varying degrees, respectively (Section \ref{sec:mstar}).
Assuming M$_{*}^{r}$ as the reference M$_{*}$ for a galaxy,
we can fit the relations between M$_{*}^{j}$ ($j$ = $i$, $z$, J, H, and K) and M$_{*}^{r}$ of the sample
in the function form below following MS14
\begin{eqnarray}
 {\rm log} (M_{*}^{j}/M_{0})&=& B_{j} {\rm log} (M_{*}^{r}/M_{0})
\end{eqnarray}
where $B_{j}$ is the slope of the linear fit line, and $M_{0}$ is the M$_{*}$ where $j$ band intersects $r$ band. 
A `robust' bi-square weighted line fit method is adopted to fit data of the LSBG sample.
The fit lines are over-plotted as red solid lines in each panel in Figure \ref{fig:mstar_discussion},
which show deviation from the line of unity in panels of $i$, $z$, J, H, and K bands,
demonstrating the problem of self-inconsistency in M$_{*}$ estimates from different bands for the same sample.
The coefficients from the fit are tabulated in Table \ref{tab:self_consistent_mass}.

\begin{table*}
\caption{Self-Consistent Stellar Masses}
\label{tab:self_consistent_mass}
\begin{center}
\begin{tabular}{lcccccccccc}
\hline
\hline
model  & $B_{i}$ & log$M_{0}^{i}$ & $B_{z}$ & log$M_{0}^{z}$ & $B_{\rm J}$ & log$M_{0}^{\rm J}$& $B_{\rm H}$ & log$M_{0}^{\rm H}$ &$B_{\rm K}$ & log$M_{0}^{\rm K}$\\
\hline
B03& 0.994&20.609& 0.994&16.132& 0.965&10.288& 0.988&13.395& 0.981&14.738\\
IP13& 0.995&17.362& 1.005& 6.132& 0.969& 6.302& 0.995&-21.17& 0.988& 0.482\\
Z09& 0.994& 9.032& 1.004&28.181& 0.968& 2.897& 0.984&-7.583& 0.983&-8.188\\
RC15(BC03)& 0.992&11.488& 0.999&-8.445& --& --& 0.981& 1.729& --& --\\
RC15(FSPS)& 0.992&13.331& 0.991&11.927& --& --& 0.976& 7.872& --& --\\
\hline
\hline
\multicolumn{11}{p{0.7\textwidth}}{\footnotesize{Notes. The coefficients are for the red solid lines in Fig. \ref{fig:mstar_discussion}
in the function form of equation (2).}}
  \end{tabular}
   \end{center} 
 \end{table*}
 
\begin{figure}
\centering
\includegraphics[width=1.0\textwidth]{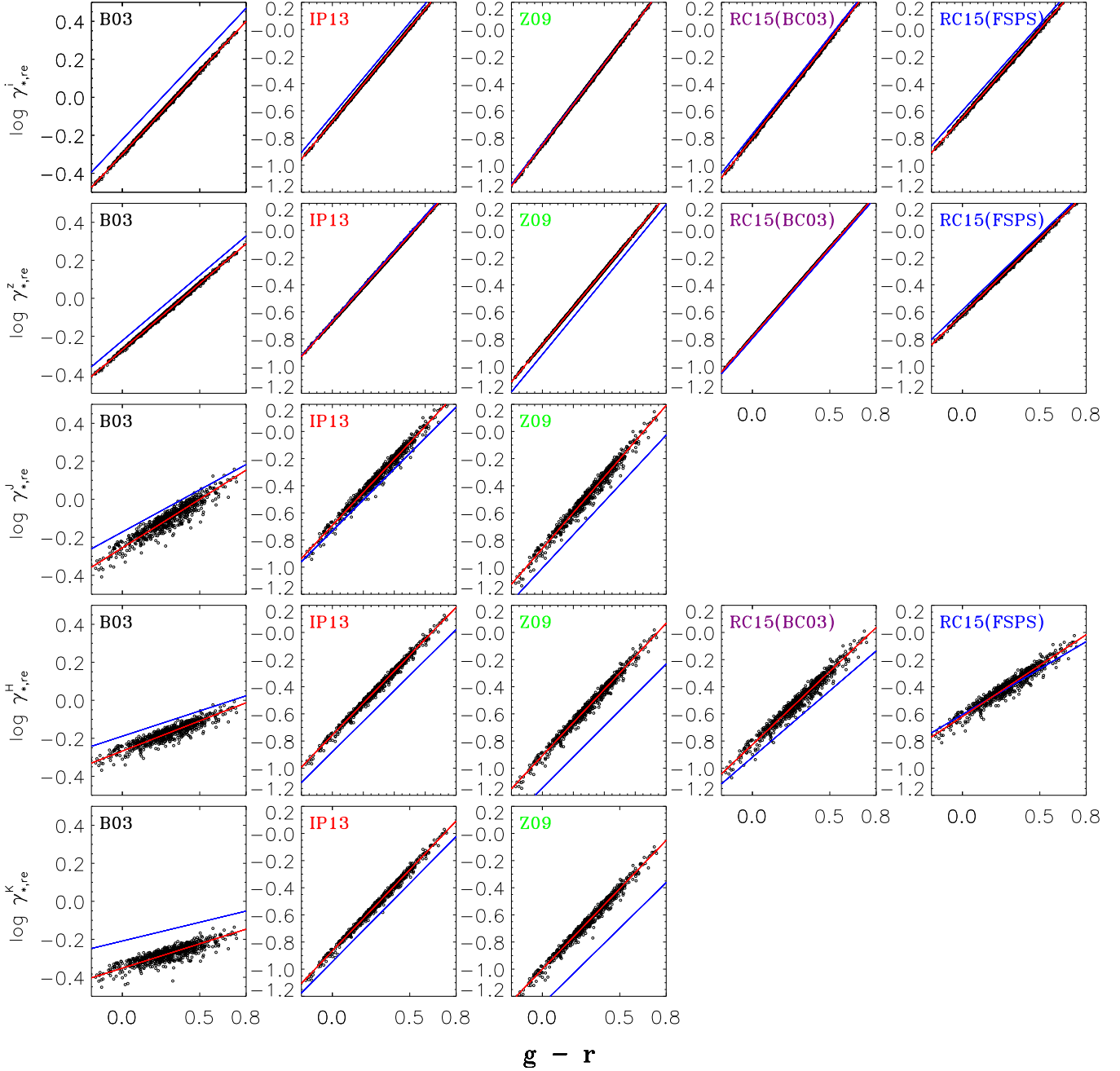} 
\caption{Renormalized stellar mass-to-light ratios ($\gamma_{*,re}^{j}, j$=$i$, $z$, J, H, and K) in logarithm as a function of $g$ - $r$ color.
Galaxies in the sample are shown as black open circles in each panel,
where the red solid line represents the fit relation between log $\gamma_{*,re}^{j}$ and $g$-$r$,
and the blue line represents the original relation for comparison. }\label{fig:self_MLCR_gr}
\end{figure}
\begin{figure}
\centering
\includegraphics[width=1.0\textwidth]{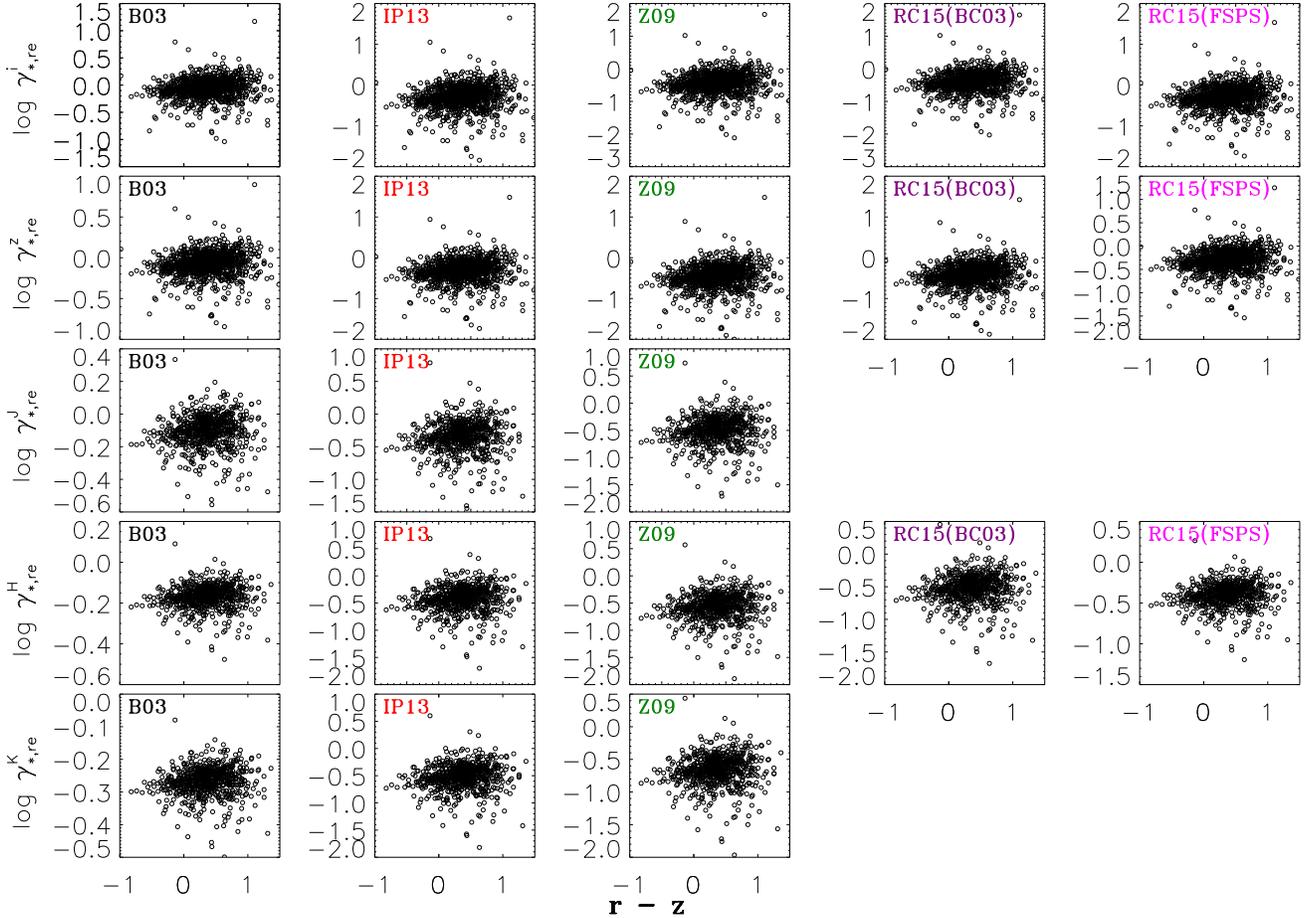}
\caption{Renormalized stellar mass-to-light ratios ($\gamma_{*,re}^{j}, j$=$i$, $z$, J, H, and K) in logrithm as a function of $r$ - $z$ color. 
The illustrations are similar to Figure \ref{fig:self_MLCR_gr}.}\label{fig:MLC_rz}
\end{figure}
\begin{figure}
\centering
\includegraphics[width=1.0\textwidth]{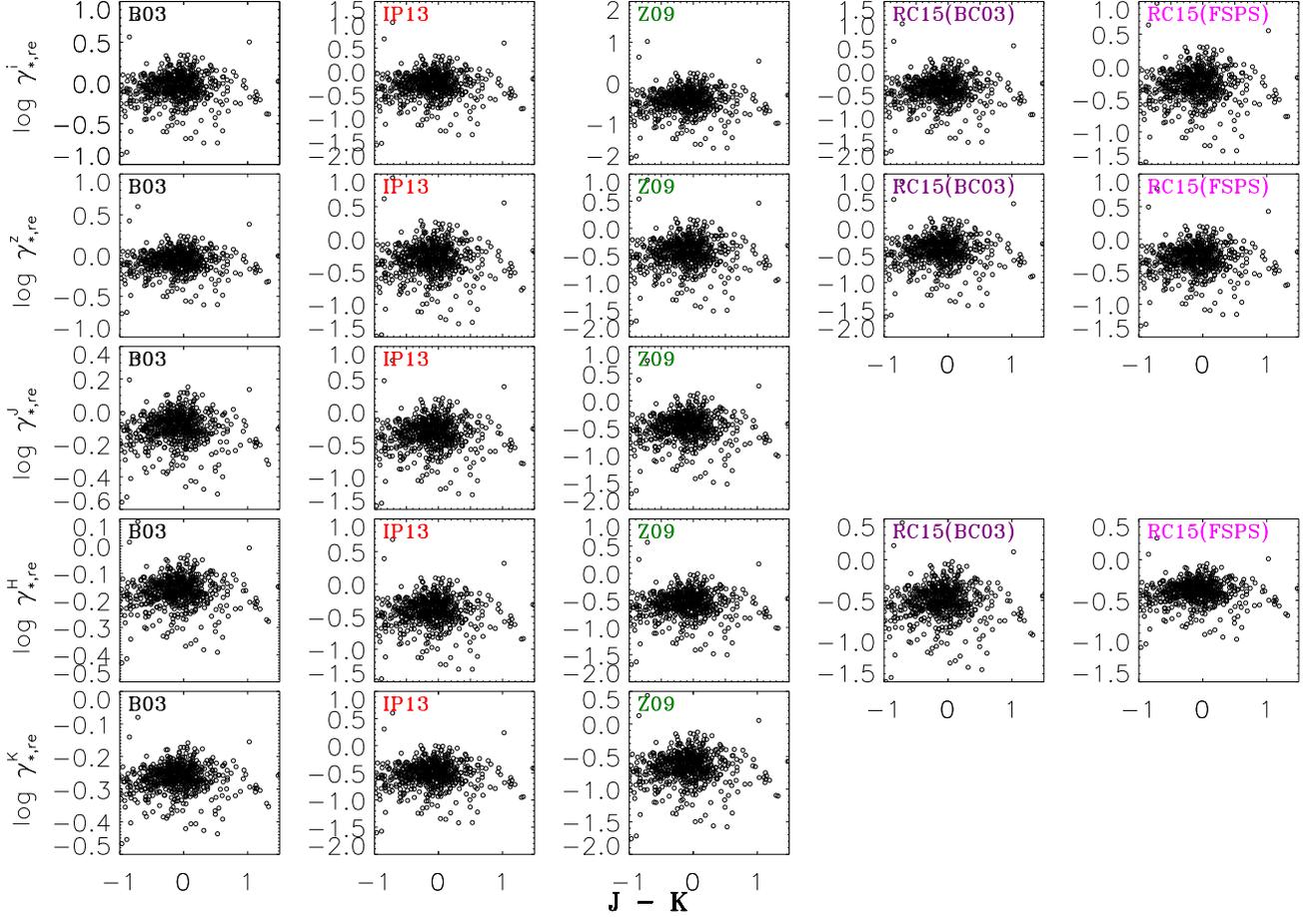}
\caption{Renormalized stellar mass-to-light ratios ($\gamma_{*,re}^{j}, j$=$i$, $z$, J, H, and K) in logrithm as a function of J - K color. 
The illustrations are similar to Figure \ref{fig:self_MLCR_gr}. }\label{fig:MLC_JK}
\end{figure}

\subsection{Re-calibrated CMLRs}\label{sec:re_cmlr}
According to the coefficients in Table ~\ref{tab:self_consistent_mass},
we renormalize M$_{*}^{j}$ ($j$=$i$, $z$, J, H, and K) to the reference mass M$_{*}^{r}$. 
Then, the renormalized M$_{*}^{j}$ (M$_{*,re}^{j}$) were divided by the luminosity in $j$ band
to generate the renormalized $\gamma_{*}^{j}$ ($\gamma_{*,re}^{j}$).
Next, the $\gamma_{*,re}^{j}$ were plotted against $g$-$r$ in Figure \ref{fig:self_MLCR_gr}.
For each panel, galaxies of the LSBG sample are shown as black open circles,
which show clear correlations between $\gamma_{*,re}^{j}$ and $g$-$r$ color. 
We then fit the relations between the $\gamma_{*,re}^{j}$ and $g$-$r$ in the function form of equation (1),
using the bi-weight line fit method. The fit line is over-plotted as red solid line in each panel in Figure \ref{fig:self_MLCR_gr},
and the blue solid line represents the original CMLRs (Table \ref{tab:cmlr1}) for comparison. 
The re-calibrated CMLRs are tabulated in Table \ref{tab:self_consistent_CMLR},
which could produce internally self-consistent M$_{*}$ estimates from different bands for the galaxy,
and this self-consistent M$_{*}$ should be highly consistent with the assumed reference mass
which is M$_{*}^{r}$ in this work.

Compared with M$_{*}^{r}$, the original B03 slightly overestimated M$_{*}$ from NIR bands (M$_{*}^{\rm NIR}$) while
the four other original CMLRs underestimated M$_{*}^{\rm NIR}$ (Table \ref{tab:mass}). 
After re-calibration, the overestimate or underestimate
are corrected correspondingly. As shown in each panel in Figure ~\ref{fig:self_MLCR_gr}, 
the re-calibrated B03 is below the original relation (blue solid line), and the four other re-calibrated relations 
of Z09, IP13, RC15(BC03), and RC15(FSPS) are all above the original relation, especially in NIR bands.
Furthermore, the original B03 require the smallest corrections,
while the original Z09 relations require the largest corrections in each band, in particular in NIR bands.
This is because Z09 is based on the prescription for the TP-AGB phase, which incorporates a larger number of TP-AGB stars.
It can greatly enhance the luminosities in the NIR but alter little the luminosities in the optical,
inevitably resulting in a lower $\gamma_{*}$ from the NIR bands than from the optical bands.

\begin{table*}\footnotesize
\caption{Re-calibrated CMLRs}
\label{tab:self_consistent_CMLR}
\begin{center}
\begin{tabular}{lcccccccccccc}
\hline
\hline
model& a$_{r}$ & b$_{r}$& a$_{i}$ & b$_{i}$ & a$_{z}$ & b$_{z}$ & a$_{\rm J}$ & b$_{\rm J}$ & a$_{\rm H}$ & b$_{\rm H}$ &a$_{\rm K}$ & b$_{\rm K}$\\
\hline
B03& -0.306& 1.097&-0.299& 0.874&-0.272& 0.699&-0.245& 0.499&-0.253& 0.283&-0.333& 0.226\\
IP13&-0.663&1.530&-0.679& 1.380&-0.674& 1.280&-0.684& 1.199&-0.742& 1.138&-0.860& 1.175\\
Z09&-0.840 & 1.654&-0.854& 1.495&-0.842& 1.374&-0.852& 1.291&-0.896& 1.178&-0.990& 1.150\\
RC15(BC03)&-0.792 & 1.629&-0.801& 1.456&-0.781& 1.308& --& --&-0.803& 1.017& --& --\\
RC15(FSPS)&-0.647 &1.497 &-0.648& 1.298&-0.619& 1.120& --& --&-0.604& 0.714& --& --\\
\hline
\hline
\multicolumn{13}{p{0.8\textwidth}}{\footnotesize{Notes. The coefficients are 
for the red solid lines in Fig. \ref{fig:self_MLCR_gr} in the function form of equation (1).}}
  \end{tabular}
   \end{center} 
 \end{table*}

\section{Discussion}\label{sec:discus}
\subsection{Secondary Color Dependence}\label{sec:discus_sub1}
$g$-$r$ acts as a primary color indicator of $\gamma_{*}$(Figure \ref{fig:self_MLCR_gr} ).
In this section, we shall examine whether the re-calibrated CMLRs based on $g$ - $r$
could be improved furthermore by including $r$-$z$ or $J$ - $K$ as a secondary color term.
Firstly, we plot $\gamma_{*,re}^{j}$ against $r$ - $z$ (Figure ~\ref{fig:MLC_rz}) or J - K (Figure ~\ref{fig:MLC_JK}) for each CMLR.
Although it appears little dependence of $\gamma_{*,re}^{j}$ on either $r$ - $z$ or J - K,
 the two colors could not be completely avoided without a further examination in quantity.
For convenience, we denote the $\gamma_{*}$ from $j$ band 
predicted by the re-calibrated CMLRs (Table \ref{tab:self_consistent_CMLR})
as $\gamma_{*,rec}^{j}$, and those predicted by the renormalized M$_{*}^{j}$ (Table \ref{tab:self_consistent_mass}) 
as $\gamma_{*,re}^{j}$ ($j$= $i$, $z$, J, H, K).
The residuals of $\gamma_{*,rec}^{j}$ from $\gamma_{*,re}^{j}$ 
 are denoted as $\Delta^{j}$ ($\Delta^{j}$=$\gamma_{*,re}^{j}$-$\gamma_{*,rec}^{j}$),
 which are in fact the difference between the data (black open circles) and 
 the re-calibrated line (red solid line) in each panel in Figure ~\ref{fig:self_MLCR_gr}.
 
\begin{figure}
\centering
\includegraphics[width=1.0\textwidth]{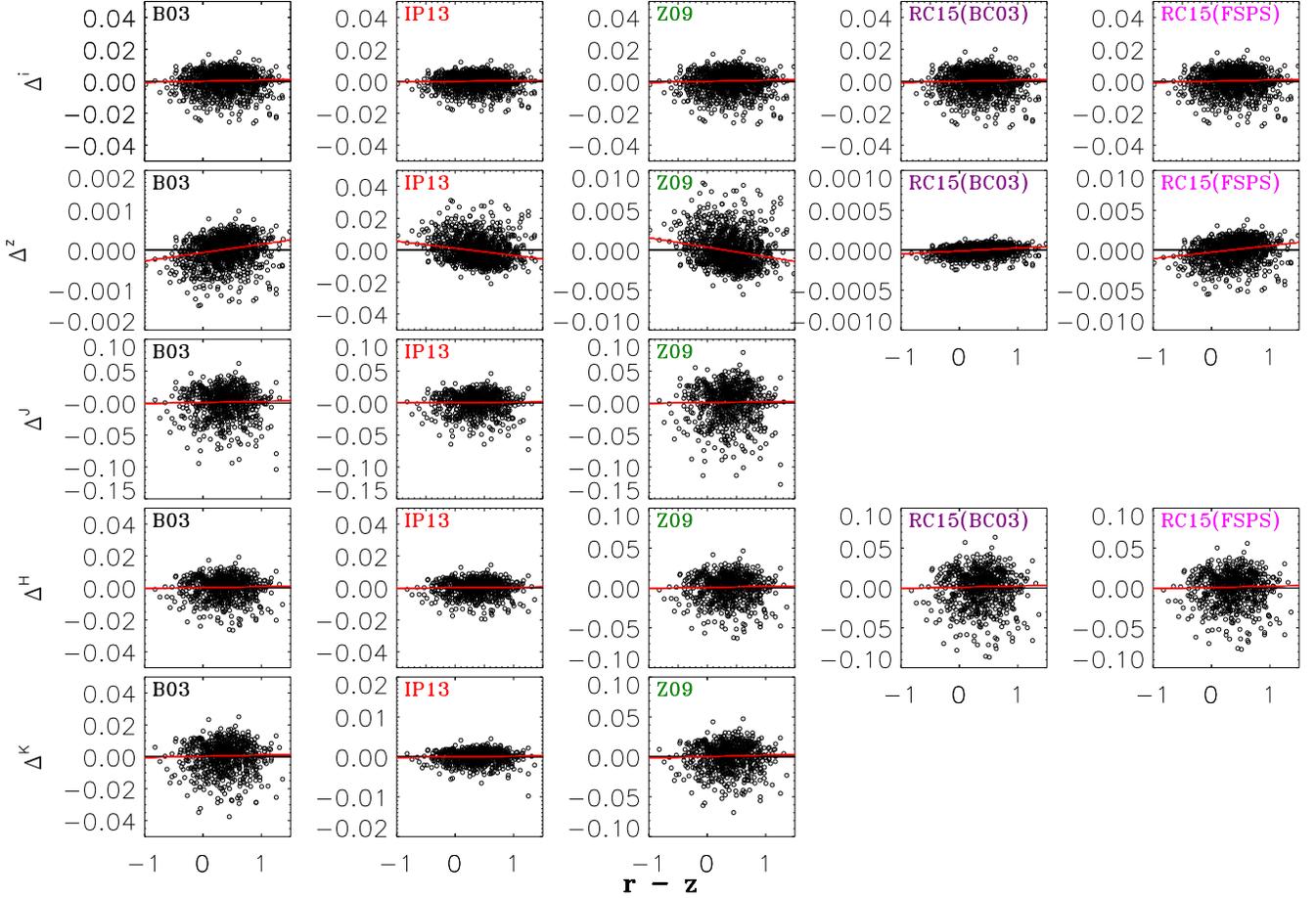}
\caption{$\Delta^{j}$ ($j$= $i$,$z$, J, H, K) as a function of $r - z$. The black line is the zero-residual line,
and the red line is the fit to the data, which nearly overlap the zero-residual line.}\label{fig:m2l_distr_rz}
\end{figure}

\begin{figure}
\centering
\includegraphics[width=1.0\textwidth]{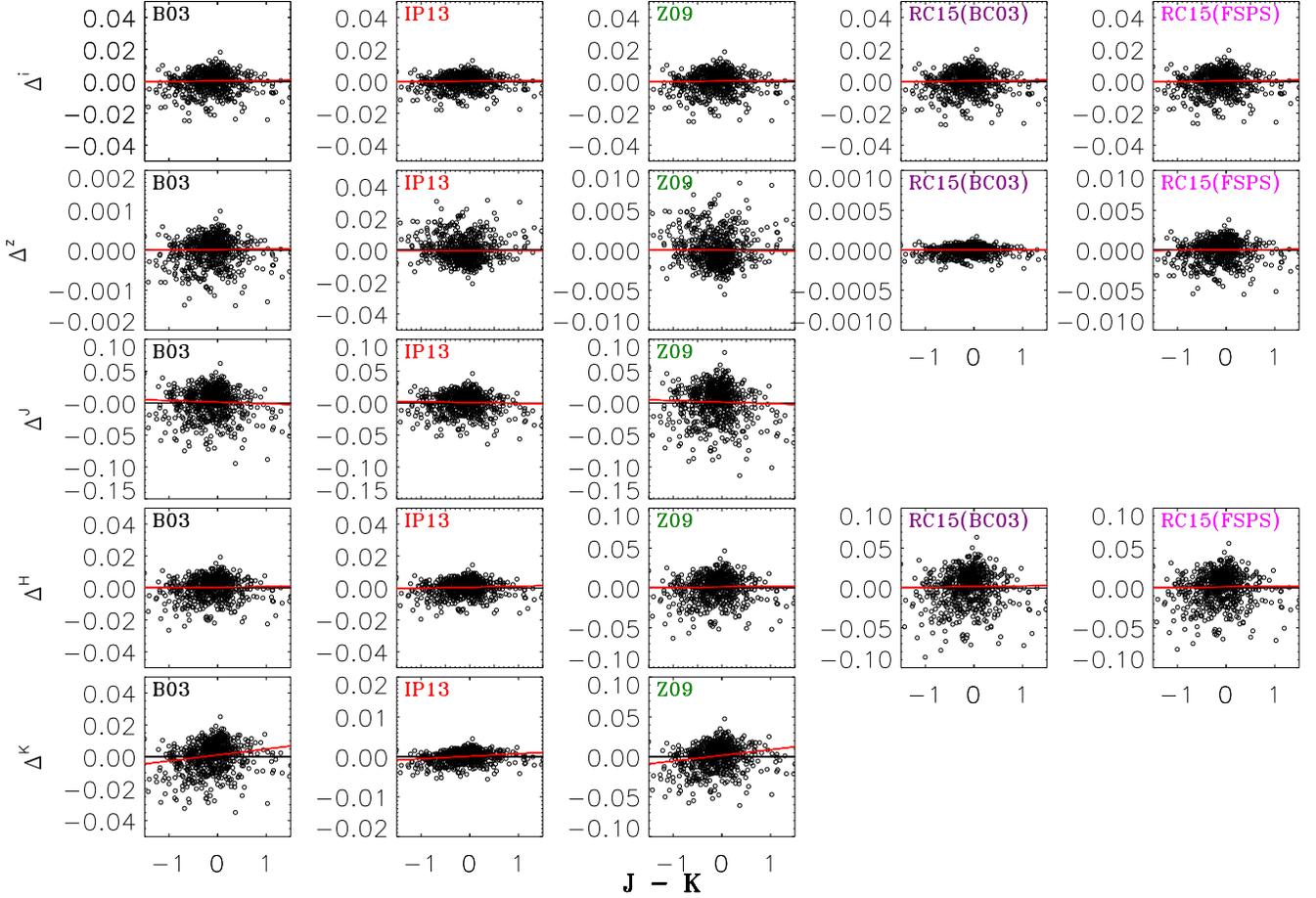}
\caption{$\Delta^{j}$ ($j$= $i$,$z$, J, H, K) as a function of J - K.The black line is the zero-residual line,
and the red line is the fit to the data, which nearly overlap the zero-residual line.}\label{fig:m2l_distr_JK}
\end{figure}

If $\Delta^{j}$ is dependent on the colors of $r$-$z$ or $r$-$z$,
 the re-calibrated CMLR based on only $g$ - $r$ color could be improved by equation (3), 
 \begin{eqnarray}
 {\rm log} \  \gamma_{*}^{j} &=& a_{j} + b_{j}\times {\rm (g - r)}+ \Delta^{j}
\end{eqnarray}
In order to check whether $\Delta^{j}$ depends on $r$-$z$ or J - K ,
we additionally plot $\Delta^{j}$ against $r$ - $z$ (Figure \ref{fig:m2l_distr_rz}) 
 or $J$ - $K$ (Figure \ref{fig:m2l_distr_JK}),
and fit a linear relation between $\Delta^{j}$ and the color in each panel (red solid line).
It shows that the fit line is almost flat and completely overlaps the zero-residual line (black line), 
implying that $\Delta^{j}$ depends little on either color of $r$ - $z$ or $J$ - $K$.
Therefore, there is no need for a secondary color term based on $r$ - $z$ or J - K ($\Delta^{j}$ in equation (3))
to improve the re-calibrated CMLRs in this work. 
This demonstrates that the variation of $\gamma_{*}$ can be well traced by the optical color but is minimized in NIR color,
which has already been proved in \citet{McGaugh2014},
where they changed the age of a solar metallicity stellar population \citet{Schombert2009} from 1 to 12 Gyr,
and the induced changes in $B$ - $V$ are 0.37 mag but only 0.03 mag in $J$ - $K$.

\begin{table*}\footnotesize
\caption{Stellar mass-to-light ratios ($\gamma_{*}$) predicted by original and re-calibrated CMLRs. }
\label{tab:cmlr2}
\begin{center}
\begin{tabular}{l|ccccc|ccccc|cc}
\hline
model&$\gamma_{0.3}^{i}$& $\gamma_{0.3}^{z}$& $\gamma_{0.3}^{J}$& $\gamma_{0.3}^{H}$& $\gamma_{0.3}^{K}$&$\gamma_{0.6}^{i}$& $\gamma_{0.6}^{z}$& $\gamma_{0.6}^{J}$& $\gamma_{0.6}^{H}$& $\gamma_{0.6}^{K}$ &$\gamma_{0.4}^{K}$ &  $\gamma_{B-V=0.6}^{K}$\\ 
\hline
\multicolumn{13}{c}{\bf{Original CMLR models}}\\
\hline
B03&1.09&0.96&0.91&0.78&0.71&1.98&1.55&1.24&0.93&0.81& 0.74 &0.73\\
IP13&0.60&0.53&0.41&0.29&0.25&1.55&1.29&0.89&0.63&0.56& 0.33 &0.41\\
Z09&0.40&0.32&0.23&0.16&0.12&1.11&0.82&0.53&0.35&0.26& 0.16 &0.21\\
RC15B&0.46&0.39&--&0.24&--&1.24&0.97&--&0.47&-- &--&--\\
RC15F&0.61&0.56&--&0.40&--&1.47&1.20&--&0.63&-- &--&--\\  
\hline
\multicolumn{13}{c}{\bf{Re-calibrated CMLR models}}\\
\hline
B03&0.92&0.87&0.79&0.67&0.53&1.68&1.40&1.13&0.84&0.63 &0.56 &0.60\\
IP13&0.54&0.51&0.47&0.40&0.31&1.41&1.24&1.09&0.89&0.71 &0.41 &0.54\\
Z09&0.39&0.37&0.34&0.29&0.23&1.11&0.96&0.85&0.67&0.51 &0.30 &0.50\\
RC15(BC03)&0.43&0.41&--&0.31&--&1.18&1.01&--&0.66&-- &--&--\\
RC15(FSPS)&0.55&0.52&--&0.40&--&1.35&1.13&--&0.68&-- &--&--\\
\hline
\multicolumn{13}{p{0.8\textwidth}}{\footnotesize{Notes. The stellar mass-to-light ratios 
predicted from different bands ($i$, $z$, J, H, K) by each CMLR before (Table ~\ref{tab:cmlr1}) 
and after re-calibration (Table ~\ref{tab:self_consistent_CMLR})  
are given at $g$ - $r$ =0.3 (the mean and the median colors of the LSBG sample)
and $g$ - $r$ = 0.6. Additionally, $\gamma_{*}$ predicted by MS14
from K band at B-V=0.6 ($\gamma_{\rm B-\rm V=0.6}^{K}$) are listed,
and for comparison, $\gamma_{*}$ predicted by our re-calibrated relations (Table ~\ref{tab:self_consistent_CMLR}) 
in Section \ref{sec:m2l} from K band at $g$-$r$=0.4 ($\gamma_{0.4}^{K}$) are also given for comparison,
since $g$-$r$=0.4 is equivalent to B - V=0.6 according to the filter transformation prescription of \citet{Smith2002}.}}
  \end{tabular}
   \end{center} 
 \end{table*}

\subsection{Error budget}\label{sec:errors}
The typical $\gamma_{*}$ uncertainties
are $\sim$0.1 ($\sim$0.2) dex in the optical (NIR) for B03,
$\sim$0.1 dex for IP13, and 0.1$\sim$0.15 dex for Z09. 
For RC15, it could be deduced (from their Figures 2 and 3 in RC15 paper) that 
the scatter in $\gamma_{*}$ from BC03 model is $\sim$0.1 dex,
but the scatter from FSPS model is not clearly available.
These typical uncertainties that are inherent in the original CMLRs
should be directly transplanted into the re-calibrated CMLRs in this work,
because the re-calibrating in this work does not change the models
on which the CMLRs are based.
For the LSBG sample, the uncertainty in $\gamma_{*}$ predicted by a CMLR
should be a combination of the inherent uncertainty in the CMLR
and the photometric error.
The uncertainty in $g$-$r$ color of the LSBG sample in this work
are  $<$  0.08 mag for 95$\%$ of the galaxies,
which would be ultimately propagated to be uncertainties of $\sim$0.08 ($\sim$ 0.03), 
$\sim$0.11 ($\sim$0.10), $\sim$0.11 ($\sim$0.10), $\sim$0.11 ($\sim$0.08), 
and $\sim$0.10 ($\sim$0.05) dex in log $\gamma_{*}$ predicted in optical (NIR) bands
by the re-calibrated relations, and almost the same values of uncertainties in log $\gamma_{*}$ 
predicted by original relations of B03, IP13, Z09, RC15(BC03), and RC15(FSPS), respectively,
Therefore, for this LSBG sample,
the total uncertainties in $\gamma_{*}$ predicted by each CMLR before or after re-calibration
are almost the same.

\subsection{$\gamma_{*}$ and M$_{*}$ from re-calibrated CMLRs}\label{sec:discus_sub2} 
In Table \ref{tab:self_consistent_CMLR}, $\gamma_{*}$ from $j$ band were estimated by each independent re-calibrated CMLR at $g$ - $r$ =0.3 
($\gamma_{0.3}^{j}$, $j$=$i$ $z$, J, H, K), which is the mean of color distribution of the sample in this work,
and $\gamma_{*}^{j}$ at $g$ - $r$=0.6 ($\gamma_{0.6}^{j}$) were also tabulated
in order to give an intuition for $\gamma_{*}^{j}$ estimates at some redder color by these re-calibrated CMLRs.
In addition, the originally predicted $\gamma_{*}^{j}$ were also listed for a comparison. 

\textbf{Apparently, B03 always gives the highest $\gamma_{*}^{j}$, 
and Z09 gives the lowest values no matter before or after re-calibration,
which is primarily due to the differences in the IMF.}
In quantity, the span in originally predicted $\gamma_{*}^{j}$ are 
$\sim$0.44, $\sim$0.48, $\sim$0.60 $\sim$0.69, and $\sim$0.77 dex at the blue color ($g$-$r$=0.3),
and are $\sim$0.25, $\sim$0.28 $\sim$0.37, $\sim$0.42, and $\sim$0.49 dex at the redder color ($g$-$r$=0.6) for $j$ = $i$, $z$, J, H, and K bands, respectively. 
In contrast, the span in $\gamma_{*}^{j}$ predicted by the re-calibrated relations has been greatly narrowed 
to $\sim$0.37, $\sim$0.37, $\sim$0.37, $\sim$0.36, and $\sim$0.36 dex at the blue color,
and to $\sim$0.18, $\sim$0.16, $\sim$0.12, $\sim$0.09, and $\sim$0.09 dex at the red color 
in the corresponding bands. So it is clear that the range in $\gamma_{*}^{j}$ by re-calibrated CMLRs is much narrower than
originally predicted, especially in the NIR bands. 
This demonstrates that the NIR luminosities are more robust than the optical luminosities
to predict the $\gamma_{*}$ of galaxies. 
It is worth noting that the uncertainties (Section \ref{sec:errors}) in $\gamma_{*}$ predicted by the original or re-calibrated relation for each CMLR
are almost the same, so these errors do not alter the comparison above.

\textbf{We can examine each re-calibrated CMLR for the internal consistency in 
M$_{*}$ from band to band. We listed the mean and median M$_{*}$ predicted by each re-calibrated CMLR 
in the right part in Table \ref{tab:mass}. It is apparent that M$_{*}^{j}$ ($j$=$g$, $i$, $z$, J, H, and K) are highly consistent with M$_{*}^{r}$ 
which is the reference stellar mass. For instance, the difference of M$_{*}^{\rm j}$ from M$_{*}^{r}$ is reduced to 0.03 dex (from the original 0.1 dex) by B03,
0.04 dex (from the original 0.3 dex) by Z09, 0.06 dex (from the original 0.27 dex) by IP13,  and 0.03 dex (from the original 0.1 - 0.2 dex) by RC15 CMLRs after re-calibration.
This demonstrates that each CMLR could produce internally self-consistent M$_{*}$ after re-calibration when it is applied in different photometric bands.}

\subsection{Comparison with MS14} \label{sec:MS14}
In the pioneering work of MS14, several CMLRs were re-calibrated in filters of V, I, K, or [3.6]
based on a sample of disk galaxies (B-V as the color indicator). 
In this work, three CMLRs that are common with MS14 were
re-calibrated, but in SDSS and NIR filters of $r$, $i$, $z$, J, H, or K based on a sample of LSBGs
($g$-$r$ as the color indicator). 
So we shall compare our re-calibrated relations with those of MS14 for the three common CMLRs (B03, IP13, and Z09)
in the common K band in this section.
 
In MS14, the $\gamma_{*}$ from K band at B-V=0.6 ($\gamma_{\rm B-\rm V=0.6}^{\rm K}$) predicted by their re-calibrated relations 
are 0.60, 0.54 and 0.50 $\rm M_{\odot}/\rm L_{\odot}$ by B03, IP13, and Z09, respectively.
In contrast, the originally predicted $\gamma_{\rm B-\rm V=0.6}^{\rm K}$ are correspondingly
0.73, 0.41, and 0.21 $\rm M_{\odot}/\rm L_{\odot}$ (the last column in Table \ref{tab:cmlr2}). 
It is apparent that the range in $\gamma_{\rm B-\rm V=0.6}^{\rm K}$ has been enormously narrowed
to 0.08 dex from the original 0.54 dex by their re-calibrations.
In order to compare with MS14, we additionally tabulate $\gamma^{\rm K}$ at $g$-$r$ = 0.4 ($\gamma_{0.4}^{\rm K}$)
predicted  by our re-calibrated relations, which are $\sim$0.57, $\sim$0.41, and $\sim$0.30 $\rm M_{\odot}/\rm L_{\odot}$ by B03, IP13, and Z09 (Table ~\ref{tab:cmlr2}),  
since $g$-$r$ = 0.4 is equivalent to B-V=0.6 according to the filter transformation prescriptions of \citet{Smith2002}.
By comparison, the originally predicted $\gamma_{0.4}^{\rm K}$ are 0.74, 0.33, and 0.16 $\rm M_{\odot}/\rm L_{\odot}$,
so the range in $\gamma_{0.4}^{\rm K}$ has been reduced to $\sim$0.28 dex from the original $\sim$0.67 dex by our re-calibrations.
However, compared with $\gamma_{\rm B-\rm V=0.6}^{\rm K}$ predicted by MS14 re-calibrated relations,
$\gamma_{0.4}^{\rm K}$ predicted by our re-calibrated relations in this work are
0.03, 0.09, 0.26 dex lower, respectively, by B03, IP13, and Z09.

In order to find out the sources of the differences,
we examined the only three different ingredients between this work and MS14, 
which are the independent procedures, the different assumptions of reference M$_{*}$, and the distinct data sets.

For the procedures, although the procedure in this work was coded to implement the same methodology as adopted by MS14, 
it is independent of and not identical with the MS14 procedure.
so we investigate the possible offset in re-calibrated relations due to
the minor differences between our and MS14 procedures, 
by repeating the exact work of MS14 on their data using our procedures.
It was found that, compared with MS14 procedures,
our procedures would drag $\gamma_{\rm B-\rm V=0.6}^{\rm K}$ down 
by 0.05, 0.01, and 0.04 dex, respectively, by B03, IP13, and Z09. 
These minor offsets in $\gamma_{\rm B-\rm V=0.6}^{\rm K}$
caused by minor differences between our and MS14 procedures
are denoted as $\Delta_{\rm pro}^{\rm K}$ for convenience (Table \ref{tab:A_tab2}).

For the assumption of reference M$_{*}$, 
we assumed M$_{*}$ estimates from SDSS $r$ band (M$_{*}^{r}$) as the reference M$_{*}$
to which M$_{*}$ estimates from other filter bands were renormalized in this work, 
while MS14 assumed M$_{*}$ from Johnson $V$ band (M$_{*}^{\rm V}$) as their reference M$_{*}$.
The different assumptions are the choices in the different filter systems (SDSS versus Johnson-Cousin), 
but it is necessary to investigate the possible offset in re-calibrated relations due to
the different choices of reference M$_{*}$ between this work and MS14 (M$_{*}^{r}$-based versus M$_{*}^{\rm V}$-based).
We present the investigation in Appendix \ref{sec:ref_mass}, 
which concludes that 
$\gamma_{\rm B-\rm V=0.6}^{\rm K}$ predicted by the M$_{*}^{r}$-based re-calibrated relations
are 0.03, 0.11, and 0.23 dex lower than those predicted by M$_{*}^{\rm V}$-based re-calibrated relations.
These major offsets in $\gamma_{\rm B-\rm V=0.6}^{\rm K}$ 
caused by the different assumptions of the reference M$_{*}$ 
are denoted as $\Delta_{\rm ref}^{\rm K}$ for convenience (Table \ref{tab:A_tab2}).

In this case, for the three common CMLRs of B03, IP13, and Z09,
the seeming differences (0.03, 0.09, 0.26 dex) between  $\gamma_{0.4}^{\rm K}$ (this work) and $\gamma_{\rm B-\rm V=0.6}^{\rm K}$ (MS14)
could be completely explained by the combination of $\Delta_{\rm ref}^{\rm K}$ (0.03, 0.11, and 0.23 dex) and 
$\Delta_{\rm pro}^{\rm K}$ (0.05, 0.01, and 0.04 dex; Table \ref{tab:A_tab2}) .
This implies that the seeming differences between our re-calibrated relations in this work
and those in MS14 in the common K band are totally caused by the systematic offsets 
due to the major differences in the assumptions of reference mass
and the minor differences in procedures between this work and MS14.
Therefore, taking into account of the different assumptions of reference mass and 
the independent procedures, our re-calibrated CMLRs based on a sample of LSBGs in this work
yield very consistent $\gamma_{*}$ in the common K band
with the re-calibrated CMLRs based on a sample of disk galaxies by MS14 .
So there is no room left for any apparent difference in the re-calibrated relations
introduced by the possible difference of our LSBG sample from the disk galaxy sample in MS14.

It is beyond the scope of this work and also difficult to evaluate which assumption of reference mass is better,
because the different assumptions are only the choices in different filter systems (SDSS versus Johnson-Cousins). 
Additionally, this work is motivated to re-calibrate each individual CMLR to give internally self-consistent M$_{*}$ for a same galaxy,
when it is applied in different bands of SDSS and NIR filters, and the internally self-consistent M$_{*}$ from any band predicted 
by each re-calibrated CMLR should be highly consistent with the reference M$_{*}$.
So, we examined the offset between different reference M$_{*}$ in Appendix,
which gives that M$_{*}^{r}$ are systematically 0.11, 0.25, and 0.33 dex lower than M$_{*}^{V}$ by B03, IP13, and Z09 (Table \ref{tab:A_tab1})
for the same sample in this work.
  
\section{Summary and Conclusions}\label{sec:conclusion}
Based on a sample of LSBGs, we examined five representative CMLRs of B03, IP13, Z09, RC15(BC03), and RC15(FSPS).
For each individual CMLR, it gives different stellar mass (M$_{*}$) estimates for the same sample, 
when it is applied in different photometric bands of SDSS optical $g$, $r$, $i$, $z$, NIR J, H, and K.
M$_{*}^{g}$ closely agree with M$_{*}^{r}$, but M$_{*}^{j}$ ($j$=$i$, $z$, J, H, K) 
all deviate from M$_{*}^{r}$, with the deviation relatively larger in NIR bands.
Assuming M$_{*}^{r}$ as a reference M$_{*}$, we re-normalized M$_{*}$ estimates from each of the other bands of $j$ (M$_{*}^{j}$) 
to the reference mass, and subsequently obtain the re-calibrated CMLR by 
fitting the relations between $g$-$r$ and $\gamma_{*}^{j}$ calculated from the re-normalized
M$_{*}^{j}$ for each original CMLR ($j$=$i$, $z$, J, H, K).
The $g$ -$r$ is the primary color indicator in the re-calibrated relations, 
which have little dependence on $r$ - $z$ or J - K. 
Each re-calibrated CMLR could produce internally self-consistent M$_{*}$ estimates for the same galaxy, 
when it is applied in different bands of $j$ ($j$=$r$, $i$, $z$, J, H, K), and the self-consistent M$_{*}$
should be ``the same as" or highly consistent with the reference mass of M$_{*}^{r}$.
Besides, the differences in original predicted $\gamma_{*}^{j}$ by the five different CMLRs 
have been largely reduced, particularly in NIR bands. 

Compared with the pioneering work of MS14, the $\gamma_{*}^{\rm K}$ predicted by the re-calibrated CMLRs
in this work are, respectively, 0.03, 0.09, 0.26 dex lower than $\gamma_{\rm B-\rm V=0.6}^{K}$ 
predicted by MS14 re-calibrations by B03, IP13, and Z09. These offsets
could be fully explained by the combination of the major systematic offsets caused by the different choices of reference mass
(0.03, 0.11, and 0.23 dex) and the minor systematic offsets caused by independent procedures (0.05, 0.01, and 0.04 dex)
between this work and MS14. This implies that, considering the major effect of different choices of reference M$_{*}$ and the minor effect
of independent procedures, the re-calibrated CMLRs in this work based on a sample of LSBGs
give very consistent $\gamma_{*}^{\rm K}$ with the re-calibrated CMLRs by MS14 at the equivalent color.
So there is no room left for any difference in the re-calibrations
caused by the possible bias of the LSB galaxy sample from the disk galaxy sample in MS14.

It is difficult to judge which choice of reference mass is better because the choices have to be made in different photometric filter systems.
However, it is necessary to give the offsets between the final self-consistent M$_{*}$ predicted by the re-calibrated relations with different
assumptions of the reference mass (M$_{*}^{r}$ versus M$_{*}^{\rm V}$). 
The M$_{*}^{r}$-based re-calibrated relations in this work (Table \ref{tab:cmlr2}) produce 
the final self-consistent M$_{*}$ which are systematically 0.11, 0.25, and 0.33 dex lower than those produced by
the M$_{*}^{\rm V}$-based re-calibrated CMLRs in MS14, by B03, IP13, and Z09.
   
\acknowledgements

The authors appreciate the anonymous referee for his/her thorough reading and constructive comments. 
D.W. would like to gratefully acknowledge China Scholarship Council (CSC) for the scholarship
which enables her as a visiting scholar to visit Professor McGaugh, S. Stacy in Astronomy Department at Case
Western Reserve University (CWRU). The work presented in this paper is fully developed and completed during her visit in CWRU.
D.W. is also supported by the National Natural Science Foundation of China (NSFC)
grant Nos. U1931109, 11733006, the Young Researcher Grant funded by National 
Astronomical Observatories, Chinese Academy of Sciences (CAS), and
the Youth Innovation Promotion Association, CAS.



%

\vspace{5mm}
\facilities{}


\software{}



\appendix
\section{Effect of reference stellar mass}\label{sec:ref_mass}
The sample in this work has photometric data in SDSS optical ($u$, $g$, $r$, $i$, $z$) and near-infrared J, H, K bands. 
In order to examine the possible effect of different assumptions of reference M$_{*}$ (M$_{*}^{r}$ versus M$_{*}^{\rm. V}$) on re-calibrated relations, 
we firstly transformed magnitudes in SDSS $g$ and $r$ filters to Johnson B and V filters
by using the transformation prescriptions of \citet{Smith2002} for the sample.
This sample now has photometric data in both SDSS filter and Johnson-Cousin B and V bands, and near-infrared J, H, and K bands. 

For the sample, M$_{*}$ from the V band (M$_{*}^{\rm V}$) or $r$ band (M$_{*}^{r}$) 
were predicted by the original CMLRs of B03, IP13 and Z09, respectively (based on B-V color). 
Comparing the distribution of M$_{*}^{\rm V}$ with that of M$_{*}^{r}$, 
M$_{*}^{r}$ is systematically 0.11, 0.25, and 0.33 dex lower than M$_{*}^{\rm V}$ by B03, IP13, and Z09 
for this same sample ($\Delta_{\rm mean}$ in Table \ref{tab:A_tab1}). 
This proves that the assumption of M$_{*}^{r}$ as reference M$_{*}$
would bias the baseline of M$_{*}$ and the re-normalized $\gamma_{*}$ in each band toward lower values,
compared with the assumption of M$_{*}^{\rm V}$ as reference M$_{*}$.

\begin{figure}
\centering  
\includegraphics[width=1.0\textwidth]{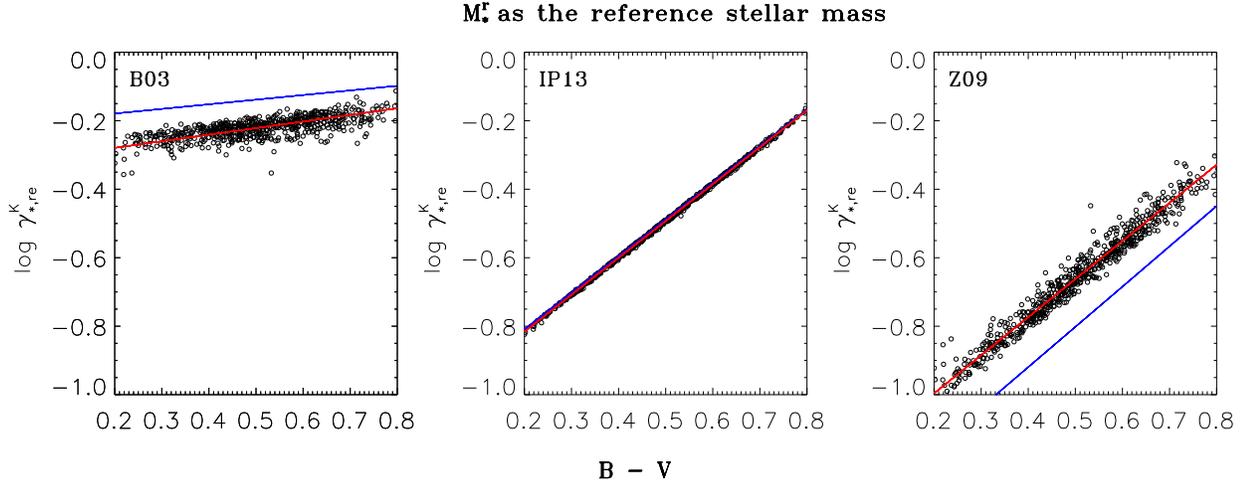}  
\caption{$\gamma_{*,re}^{\rm K}$ from the stellar mass renormalized to M$_{*}^{r}$ are plotted against B-V color.
For each panel, the black circles are galaxies of the LSBG sample, and the red solid line is the fit to the data (re-calibrated CMLR),
and the blue line is the original CMLR.}  
\label{fig:fig1} 
\end{figure}

\begin{figure}
\centering  
\includegraphics[width=1.0\textwidth]{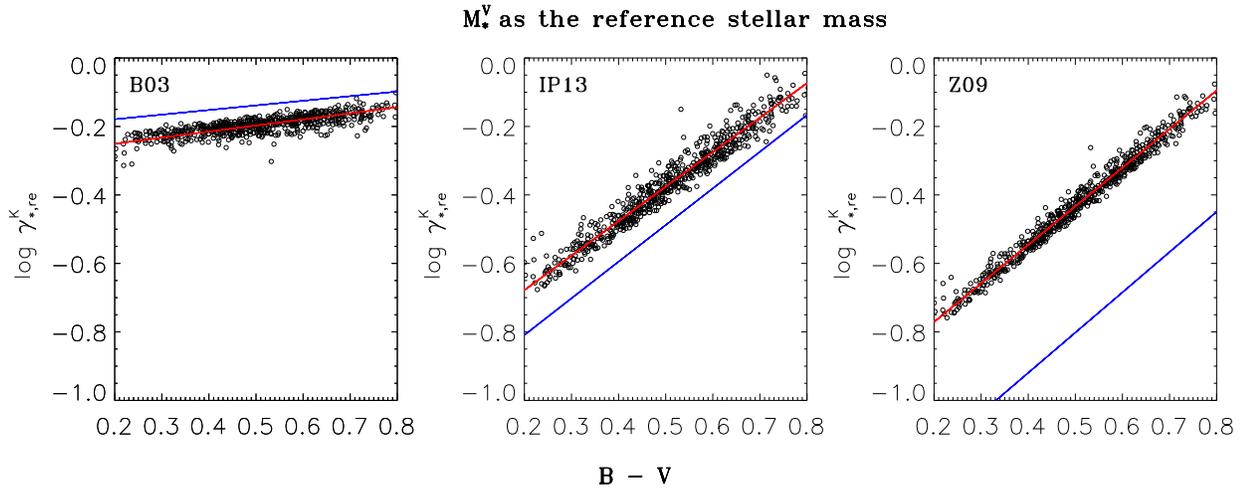}  
\caption{$\gamma_{*,re}^{\rm K}$ from the stellar mass renormalized to M$_{*}^{\rm V}$ are plotted against B-V color.
For each panel, the black circles are galaxies of the LSBG sample, and the red solid line is the fit to the data (re-calibrated CMLR),
and the blue line is the original CMLR. }  
\label{fig:fig2} 
\end{figure}

In analogy to Section \ref{sec:m2l}, we furthermore re-calibrated each of the three CMLRs (B03, IP13, Z09) in K band (on B-V color) based on the sample of LSBGs, 
assuming M$_{*}^{V}$ or M$_{*}^{r}$ as reference M$_{*}$, respectively. 
More specifically, for each of the three CMLRs (B03, IP13, Z09), we first re-normalized M$_{*}$ estimates from K band (M$_{*}^{\rm K}$) to the reference mass of M$_{*}^{r}$,
then divided the re-normalized M$_{*}^{\rm K}$ by the K-band luminosity to obtain the re-normalized $\gamma_{*}^{\rm K}$ ($\gamma_{*,re}^{K}$),
and ultimately fit the relations between $\gamma_{*,re}^{\rm K}$ and B-V color to obtain the re-calibrated relations \textbf{in K band}, which is denoted as 
M$_{*}^{r}$-based re-calibrated relations for convenience. Similarly, we obtained M$_{*}^{\rm V}$-based re-calibrated relations in $m$ band by assuming
the reference mass of M$_{*}^{\rm V}$. These two sets of re-calibrated CMLRs in K band are shown in Figures \ref{fig:fig1} - \ref{fig:fig2},
where it clearly shows that, compared with the M$_{*}^{\rm V}$-based re-calibrations (red solid line in Figures \ref{fig:fig2}),
the M$_{*}^{r}$-based re-calibrated relations (red solid line in Figures \ref{fig:fig1}) obviously dragged down $\gamma_{*,re}$ in each panel for each CMLR,
in particular for IP13 and Z09, since the two figures share the same y-axis range and scale for convenient comparison.

In Table \ref{tab:A_tab2}, we listed $\gamma_{*}$ in K band at B-V =0.6 predicted by M$_{*}^{\rm V}$-based 
($\gamma_{B-V=0.6}^{K}$(M$_{*}^{\rm V}$) ) and M$_{*}^{r}$-based ($\gamma_{B-V=0.6}^{K}$(M$_{*}^{r}$)) re-calibrated CMLRs, 
and $\gamma_{B-V=0.6}^{K}$ predicted by MS14 ($\gamma_{B-V=0.6}^{K}$(MS14)) for a comparison.
It is apparent that $\gamma_{B-V=0.6}^{K}$(M$_{*}^{V}$) are very consistent with $\gamma_{B-V=0.6}^{K}$(MS14) which also assumed M$_{*}^{V}$ as reference M$_{*}$,
and the small differences between the two are caused by the minor difference in the procedures between this work and MS14 ($\Delta_{\rm pro}$ in Table \ref{tab:A_tab2} and
already discussed in Section \ref{sec:MS14}). This implies that assuming M$_{*}$ from the same V band as the reference M$_{*}$,
the re-calibrated CMLRs in K band (on B-V color) based on the sample of LSBGs
give very consistent $\gamma_{B-V=0.6}^{K}$ with those given by MS14 based on a sample of disk galaxies,
which further indicates that there appears no apparent bias in $\gamma_{B-V=0.6}^{K}$ that was introduced by differences in samples between this work and MS14.

Compared with $\gamma_{B-V=0.6}^{K}$(M$_{*}^{V}$), $\gamma_{B-V=0.6}^{K}$(M$_{*}^{r}$)
are 0.03, 0.11, and 0.25 dex lower by B03, IP13, and Z09. These offsets should be only caused by the difference in the assumption of reference M$_{*}$ (M$_{*}^{r}$ or M$_{*}^{V}$),
so they were denoted as $\Delta_{\rm ref}$ in Table \ref{tab:A_tab2},
since these two sets of re-calibrated relations only differ in the assumption of reference M$_{*}$.
This implies that different assumptions of reference  M$_{*}$ (M$_{*}^{r}$ or M$_{*}^{V}$)
would cause evident offsets in $\gamma_{B-V=0.6}^{K}$ values predicted by re-calibrated CMLRs. 
Quantitatively, M$_{*}^{r}$-based re-calibrated CMLRs give  $\gamma_{B-V=0.6}^{K}$
that are 0.03, 0.11, and 0.25 dex systematically lower than those given by 
M$_{*}^{V}$-based re-calibrated CMLRs (Table \ref{tab:A_tab2}). 

For the difference between $\gamma_{0.4}^{K}$ and $\gamma_{B-V=0.6}^{K}$ discussed in Section \ref{sec:MS14}, 
$\gamma_{0.4}^{K}$ are 0.57, 0.42, 0.30, and $\gamma_{B-V=0.6}^{K}$ 
are 0.60, 0.50, 0.54 by B03, IP13 and Z09 after re-calibration, 
and the difference between the two are 0.02, 0.08, 0.26 dex, by B03, IP13, and Z09 (Table \ref{tab:cmlr2}).
Numerically, the difference could be fully explained by the major offsets caused by the different assumptions of reference M$_{*}$ ($\Delta_{\rm ref} \sim$  0.03, 0.11, and 0.23 dex) and 
the minor offsets due to differences in procedures between this work and MS14 ($\Delta_{\rm pro} \sim <$ 0.05 dex) (Table \ref{tab:A_tab2}).
This explanation should be plausible because $\gamma_{B-V=0.6}^{K}$ was predicted by MS14, which assumed M$_{*}^{V}$ as the reference mass,
while $\gamma_{0.4}^{K}$ was predicted by our re-calibrated relations (Table \ref{tab:cmlr1}), which assumed M$_{*}^{r}$ as the reference mass.
So considering the different assumptions of the reference M$_{*}$ and the minor difference in procedures between this work and MS14,
the re-calibrated CMLRs (B03, IP13, Z09) in SDSS filters in this work (Table \ref{tab:cmlr1}) are fundamentally consistent in $\gamma_{*}^{K}$ at the equivalent colors
with those in Johnson-Cousin filters by MS14. 

In a brief, the analysis in this Appendix demonstrate that,compared with an assumption of M$_{*}^{V}$  as the reference mass (MS14),
the assumption of M$_{*}^{r}$ as the reference mass (Section \ref{sec:re_cmlr} ) would lower the self-consistent (also reference) M$_{*}$ down
by 0.11, 0.25, and 0.33 dex, and lower $\gamma_{*}^{K}$ down by 0.03, 0.11, and 0.25 dex by B03, IP13, and Z09.

\begin{table}
\caption{Distribution (mean, $\sigma$) of M$_{*}$ from SDSS $r$ or Johnson V bands of the LSBG sample.}
\label{tab:A_tab1}
\begin{center}
\begin{tabular}{c|cc|cc|c}
\hline
\hline
model  & mean(M$_{*}^{r}$) & $\sigma$(M$_{*}^{r}$) &mean(M$_{*}^{\rm V}$) & $\sigma$(M$_{*}^{\rm V}$)&$\Delta_{\rm mean}$\\
\hline
B03 & 8.65 & 0.81 &8.76 & 0.75 &0.11\\
IP13 & 8.41 &0.85 &8.66 &0.77 &0.25\\
Z09 & 8.27 & 0.87 & 8.60 & 0.79 &0.33\\
\hline
\multicolumn{6}{p{0.5\textwidth}}{Notes. The values are all in logarithm. M$_{*}^{r}$ is estimated from 
the $r$-band luminosities with the $g$-$r$ as the indicator color of $\gamma_{*}^{r}$.
M$_{*}^{\rm V}$ is estimated from the V-band luminosities with the B-V as the indicator color of $\gamma_{*}^{\rm V}$.
$\Delta_{\rm mean}$ is the difference of the mean value of M$_{*}^{r}$ distribution from that of M$_{*}^{\rm V}$ distribution.}
  \end{tabular}
   \end{center} 
 \end{table}

\begin{table}
\caption{$\gamma_{*}^{\rm K}$ predicted by M$_{*}^{r}$-based or M$_{*}^{\rm V}$-based re-calibrated CMLRs at B-V = 0.6 in this work,
and those from M$_{*}^{\rm V}$-based re-calibrations in MS14. }
\label{tab:A_tab2}
\begin{center}
\begin{tabular}{lccc|cc}
\hline
\hline
model&$\gamma_{B-V=0.6}^{\rm K}$(M$_{*}^{r}$) & $\gamma_{B-V=0.6}^{\rm K}$(M$_{*}^{\rm V}$) & $\gamma_{0.6}^{\rm K}$(MS14)&$\Delta_{\rm ref}^{\rm K}$ &$\Delta_{\rm pro}^{\rm K}$\\
            
\hline
 B03 & 0.62 & 0.66 & 0.60 & 0.03  & 0.05 \\
 IP13 & 0.41 & 0.53 & 0.50& 0.11  &0.01  \\
 Z09 & 0.28 & 0.48 & 0.54 & 0.25  & 0.04 \\
  \hline
\hline
\multicolumn{6}{p{0.6\textwidth}}{\footnotesize{Notes.
$\Delta_{\rm ref}^{\rm K}$ is the difference between $\gamma_{0.6}^{\rm K}$(M$_{*}^{r}$) and $\gamma_{0.6}^{\rm K}$(M$_{*}^{\rm V}$).
$\Delta_{\rm pro}^{\rm K}$ is the systematic bias in $\gamma_{0.6}^{\rm K}$ caused by the minor difference in procedures between 
this work and MS14.}}

 \end{tabular}
   \end{center} 
 \end{table}

\clearpage



\end{document}